\documentclass[12pt]{article}
\pdfoutput=1

\usepackage{hyperref,feynmp}
\usepackage{amsmath,amssymb,amssymb,graphicx} 
\usepackage{cite}
\usepackage{slashed}

\newcommand{\beq}{\begin{eqnarray}}
\newcommand{\eeq}{\end{eqnarray}}

\numberwithin{equation}{section}

\newcommand{\centeron}[2]{{\setbox0=\hbox{#1}\setbox1=\hbox{#2}\ifdim
\wd1>\wd0\kern.5\wd1\kern-.5\wd0\fi \copy0
\kern-.5\wd0\kern-.5\wd1\copy1\ifdim\wd0>\wd1
                                   \kern.5\wd0\kern-.5\wd1\fi}}
\newcommand{\ltap}{\>\centeron{\raise.35ex\hbox{$<$}}
                           {\lower.65ex\hbox{$\sim$}}\>}
\newcommand{\gtap}{\>\centeron{\raise.35ex\hbox{$>$}}
                           {\lower.65ex\hbox{$\sim$}}\>}
\newcommand{\gsim}{\mathrel{\gtap}}
\newcommand{\lsim}{\mathrel{\ltap}}

\newcommand\ZZ{\hbox{\zfont Z\kern-.4emZ}}
\font\zfont = cmss10 

\newcommand{\drawsquare}[2]{\hbox{%
\rule{#2pt}{#1pt}\hskip-#2pt
\rule{#1pt}{#2pt}\hskip-#1pt
\rule[#1pt]{#1pt}{#2pt}}\rule[#1pt]{#2pt}{#2pt}\hskip-#2pt
\rule{#2pt}{#1pt}}

\newcommand{\fund}{\ensuremath{\drawsquare{6.5}{0.4}}}
\newcommand{\afund}{\ensuremath{\overline{\fund}}}

\newcommand{\symm}{\ensuremath{\drawsquare{6.5}{0.4}\hskip-0.4pt%
        \drawsquare{6.5}{0.4}}}

\newcommand{\sing}{\ensuremath{\mathbf{1}}}

\def\lpar#1#2#3#4{\rlap{\raise#3\hbox{$\hskip#4#1\left\{\mbox{\phantom{\rule[0mm]{0mm}{#2}}}\right.$}}}
\def\rpar#1#2#3#4{\rlap{\raise#3\hbox{$\hskip#4\left\}#1\mbox{\phantom{\rule[0mm]{0mm}{#2}}}\right.$}}}

\begin{document}
\setlength{\unitlength}{1mm}
\begin{titlepage}

\vskip.5cm
\begin{center}
{\Large \bf MFV SUSY: A Natural Theory
 for R-parity Violation}


\vskip.1cm
\end{center}
\vskip0.2cm

\begin{center}
{
{Csaba Cs\'aki}, {Yuval Grossman},
{\rm and}
{Ben Heidenreich}}
\end{center}
\vskip 8pt

\begin{center}
{\it  Department of Physics, LEPP, 
Cornell University, Ithaca, NY 14853} \\
\vspace*{0.1cm}
 
\vspace*{0.3cm}
{\tt  csaki,yg73,bjh77@cornell.edu}
\end{center}

\vglue 0.3truecm

\begin{abstract}
\vskip 3pt
\noindent

We present an alternative approach to low-energy
supersymmetry. Instead of imposing R-parity    we apply the minimal
flavor violation (MFV) hypothesis to the R-parity violating MSSM.  In
this framework, which we call MFV SUSY, squarks can be light and the
proton long lived without producing missing energy signals at the
LHC. Our approach differs from that of Nikolidakis and Smith in that
we impose holomorphy on the MFV spurions. The resulting model is
highly constrained and R-parity emerges as an accidental approximate
symmetry of the low-energy Lagrangian.  The size of the small R-parity
violating terms is determined by the flavor parameters, and in the
absence of neutrino masses there is only one renormalizable R-parity
violating interaction: the baryon-number violating
$\bar{u}\bar{d}\bar{d}$ superpotential term. Low energy observables
(proton decay, dinucleon decay and $n-\bar{n}$ oscillation) pose only
mild constraints on the parameter space. 
LHC phenomenology will depend on whether the LSP is a squark,
neutralino, chargino or slepton. If the LSP is a squark it will have
prompt decays, explaining the non-observation of events with missing
transverse energy at the LHC.

\end{abstract}

\end{titlepage}


\section{Introduction}
\label{sec:intro}
\setcounter{equation}{0}
\setcounter{footnote}{0}

Supersymmetric extensions of the standard model do not automatically posses the requisite global symmetries of the standard model: baryon and lepton number violation can be mediated by squark and gaugino exchange, and flavor-non-universal soft breaking terms can mediate flavor-changing neutral currents (FCNCs). In order to remove baryon and lepton number violating processes one usually assumes the additional presence of R-parity, while to remove FCNCs one usually assumes flavor universality (possibly at a high scale). R-parity has very important consequences for the phenomenology of the MSSM: it renders the lightest superpartner stable, forces superpartners to be pair-produced, and implies that (when produced) superparticles will always decay to the LSP, which will escape the detector, resulting in events with  large missing energy.

R-parity is clearly not necessary~\cite{HallSuzuki,RossValle,Tao,Herbi,Bhattacharyya,Mohapatra}: very small R-parity violating terms can be added to the supersymmetric Lagrangian, fundamentally changing the phenomenology of the model without conflicting with any current experimental bound (for an excellent review see~\cite{Barbier:2004ez}).
The introduction of R-parity is therefore linked to the idea of naturalness: if R-parity were not imposed, many dimensionless couplings in the superpotential would have to be extremely small in order to ensure a sufficiently long-lived proton.

LHC data, however, is beginning to place severe constraints on the R-parity conserving MSSM, ruling out squark masses up to about $1\mathrm{\ TeV}$ in some scenarios, due to the absence of the expected missing transverse energy events~\cite{Chatrchyan:2011zy,AtlasSUSYsearch}. Increasing the scale of supersymmetry breaking leads to increasingly large radiative corrections to the Higgs mass,
suggesting that low-scale supersymmetry with R-parity may not be the
correct solution to the hierarchy problem. In light of this it is
natural to consider R-parity violation, which allows the LSP to decay
promptly, and thus evades searches based on missing transverse energy
or displaced vertices. However, besides naturalness, such an
undertaking suffers from a proliferation of undetermined couplings, making
it very difficult to constrain the theory from experimental data.

Here, we consider an alternate approach to low-energy supersymmetry. Instead of assuming R-parity, we only impose the minimal flavor violation hypothesis on the theory~\cite{Sekhar,Lisa,MFVBuras,MFVEFT}, positing that the non-abelian flavor symmetries are only broken by the holomorphic spurions corresponding to the Yukawa couplings.\footnote{While the most general flavor symmetry, U$(3)^5$, is not semi-simple, the abelian U$(1)^5$ component contains R-parity, and would imply the complete absence of lepton- and baryon-number violating operators. In our spurion analysis, we only impose the nonabelian SU$(3)^5$ component.}  As a consequence, all R-parity
violating operators will be suppressed by Yukawa couplings and CKM
factors, and the smallness of the R-parity violating terms is explained in terms of the smallness of the flavor parameters.  We find that this assumption is sufficient to naturally avoid
present bounds on baryon- and lepton-number violation, while automatically suppressing FCNCs as in any MFV model. Thus, we
are able to replace two independent ad-hoc assumptions, those of
R-parity and flavor universality, with the single assumption of
minimal flavor violation. R-parity then emerges as an approximate accidental symmetry of the low-energy Lagrangian, where the R-parity breaking terms are determined by the flavor sector.

We will argue that the simplest form of this model is viable with
natural ${\cal O}(1)$ coefficients for all operators and low, $\sim100
- 300\;\mathrm{GeV}$, superpartner masses. This provides a natural
alternative framework for studying supersymmetric extensions of the
standard model. While the R-parity violating couplings are
sufficiently small to prevent proton decay, they are sufficiently
large to make the LSP decay promptly.
The phenomenology is distinctive, and depends on only a relatively
small number of unknown $\mathcal{O}(1)$ parameters, in contrast to
the generic R-parity violating MSSM.

The idea that minimal flavor violation can replace R-parity was originally explored in an important paper by Nikolidakis and
Smith a few years ago~\cite{Nikolidakis:2007fc} (see
also~\cite{Smith}). Our approach differs from theirs in that we take
the spurions to be holomorphic, which is necessary since they appear in
the superpotential as Yukawa couplings, and should be thought of as VEVs
of chiral superfields. Thus, $Y^\dag$ cannot appear in the
superpotential, nor in soft-breaking $A$-terms,\footnote{Nonholomorphic corrections to the $A$-terms are possible. However, these corrections are subleading, as explored in Appendix~\ref{app:SUSYbreaking}. In addition, bilinear corrections to the superpotential can be generated nonholomorphically at the scale $m_{\rm soft}$.} which,
combined with the MFV hypothesis, severely constrains the form of
these terms.

We will show that in the absence of neutrino
masses there is no holomorphic invariant violating lepton number, and
there is only a single renormalizable term violating baryon number,
the $\bar{u}\bar{d}\bar{d}$ term in the superpotential. Furthermore,
an unbroken $\mathbb{Z}_3^L$ subgroup of U$(1)_L$ --- a necessary
consequence of MFV --- ensures that the first non-holomorphic
(K\"{a}hler) corrections violating lepton number appear at dimension
eight, and are very strongly suppressed for even a moderately high
cutoff scale. Thus, in the limit of vanishing neutrino masses the
proton will be effectively stable. The constraints from
$n-\bar{n}$ oscillations are easily satisfied, while those from dinucleon decay place a mild upper
bound on $\tan\beta$ for light squark masses.

Majorana neutrino masses require additional holomorphic spurions
charged under $\mathbb{Z}_3^L$, and we find that once they are
incorporated into the model through the seesaw mechanism, current
bounds on proton decay will impose interesting, though not too
onerous, constraints on the right-handed neutrino sector. Other
methods of neutrino mass generation should also be constrained by
proton stability.

The phenomenology of such models is largely determined by the choice
of the LSP. If it is a squark, it can decay directly via the
baryon number violating $\bar{u}\bar{d}\bar{d}$ vertex, which yields a
lifetime short enough for these decays to be prompt. If a sparticle
other than a squark is the LSP (such as a neutralino, chargino or
slepton) then the decays will involve more particles in the final
state and the lifetime will increase, potentially leading to displaced
vertices, and in some cases also to missing energy via neutrinos and
tops in the final state.
 
The paper is organized as follows. In \S\ref{sec:setupmassless} we introduce the
MFV SUSY framework and list possible superpotential terms, neglecting
neutrino masses. In \S\ref{sec:BNVvertex}, we focus on the most interesting of these terms, a baryon number violating vertex.
In \S\ref{sec:consmassless}, we discuss constraints
arising from $n-\bar{n}$ oscillations and dinucleon decay induced by this vertex. In
\S\ref{sec:setupneutrinos}, we modify the model to incorporate
neutrino masses, focusing on the seesaw mechanism, and list the
relevant operators, VEVs, and mixings. In \S\ref{sec:consneutrinos},
we discuss constraints on the right-handed neutrino sector arising
from bounds on proton decay. In \S\ref{sec:consequences} we estimate
the LSP lifetime and comment on LHC signals/constraints. We
conclude in \S\ref{sec:conclusions}. In a collection of appendices, we classify all possible holomorphic superpotential terms, discuss nonholomorphic corrections from supersymmetry breaking, argue that diagrams other than those considered in the main text will be subdominant for the processes of interest, and show that higher-dimensional operators will not affect our conclusions for a sufficiently high cutoff.



\section{MFV SUSY without neutrino masses}\label{sec:setupmassless}

\begin{table}[t]
\begin{center}
\begin{tabular}{c|ccc}
& SU$(3)_C$ & SU$(2)_L$ & U$(1)_Y$ \\
\hline
$Q$ & \fund & \fund & $1/6$ \\
$\bar{u}$ & \afund & \sing & $-2/3$ \\
$\bar{d}$ & \afund & \sing & $1/3$ \\
$L$ & \sing & \fund & $-1/2$ \\
$\bar{e}$ & \sing & \sing & $1$ \\
\hline
$H_u$ & \sing & \fund & $1/2$ \\
$H_d$ & \sing & \fund & $-1/2$ \\
\end{tabular}
\end{center}
\caption{The MSSM fields and their representations under the SM gauge
group.\label{tab:mssm}}
\end{table}

We first consider the limit of vanishing neutrino masses (we introduce
them  in~\S\ref{sec:setupneutrinos}).  The MSSM consists of the
standard model (SM) gauge group $\mathrm{SU}(3)_C\times
\mathrm{SU}(2)_L \times \mathrm{U}(1)_Y$, together with the usual chiral
superfields as shown in Table~\ref{tab:mssm}.  The matter fields $Q,
\bar{u}, \bar{d}, L,$ and $\bar{e}$ are flavored, and come in three
generations. The superpotential
\begin{equation} \label{eqn:RPCrenormW}
W = \mu H_u H_d + Y_e L H_d \bar{e} + Y_u Q H_u  \bar{u} + Y_d Q H_d  \bar{d} \, ,
\end{equation}
is necessary to generate the SM fermion masses and charged higgsino masses. The additional (renormalizable) superpotential terms allowed by gauge invariance are
\begin{equation}
W'=\lambda LL\bar{e}+\lambda' QL\bar{d} +\lambda'' \bar{u}\bar{d}\bar{d} + \bar{\mu} LH_u \, .
\label{eq:RPVrenormW}
\end{equation}
These superpotential terms violate lepton and baryon number, and
therefore should be absent or very small. The traditional approach is
to impose a $\mathbb{Z}_2$ symmetry, called matter parity, under which
the matter fields $Q, \bar{u}, \bar{d}, L,$ and $\bar{e}$ are odd and
the Higgs fields $H_u$ and $H_d$ are even. This $\mathbb{Z}_2$
symmetry forbids all unwanted superpotential terms in $W'$, leaving
only those in~(\ref{eqn:RPCrenormW}). A combination of matter parity
with a discrete subgroup of the Lorentz group gives R-parity, under
which all SM fields are even and superpartners odd. 

The imposition of R-parity is not the only ad-hoc assumption needed to
make the MSSM phenomenologically acceptable. Soft terms needed to
break supersymmetry and mass-up the superpartners generically induce
large flavor-changing neutral currents. In order to reduce FCNCs, one usually imposes flavor universality: i.e.\ the assumption that at some scale all soft breaking masses are flavor universal and the $A$-terms
are proportional to the corresponding Yukawa couplings.

Our approach will be to replace
these two ad-hoc assumptions with the single assumption of Minimal
Flavor Violation (MFV). MFV is based on the observation that apart
from the $\mu$ term, most of the terms in the superpotential
(\ref{eqn:RPCrenormW}) are small due to the smallness of the Yukawa
couplings. It is then natural to analyze the spurious symmetries
preserved by the $\mu$-term but broken by the Yukawa couplings, which
are given in Table~\ref{tab:flavor}.
Excepting $U(1)_{B-L}$ and a $U(1)^2$ subgroup of $\mathrm{SU}(3)_L\times \mathrm{SU}(3)_e$ representing intergenerational lepton number differences, the Yukawa couplings are charged under all of these symmetries, which are therefore broken by the superpotential.

\begin{table}[t]
\begin{center}
\begin{tabular}{c|ccccc|cc}
& SU$(3)_Q$ & SU$(3)_u$ & SU$(3)_d$ & SU$(3)_L$ & SU$(3)_e$ & U$(1)_{B-L}$ & U$(1)_H$ \\
\hline
$Q$ & \afund & \sing & \sing & \sing & \sing & $1/3$ & $0$\\
$\bar{u}$ & \sing & \fund & \sing & \sing & \sing & $-1/3$ & $0$\\
$\bar{d}$ & \sing & \sing & \fund & \sing & \sing & $-1/3$ & $0$\\
$L$ &  \sing & \sing & \sing & \afund & \sing & $-1$ & $0$\\
$\bar{e}$ & \sing & \sing & \sing & \sing & \fund & $1$ & $0$\\
$H_u$ & \sing & \sing & \sing & \sing & \sing & $0$ & $1$\\
$H_d$ & \sing & \sing & \sing & \sing & \sing & $0$ & $-1$\\
\hline
$Y_u$ & \fund & \afund & \sing & \sing & \sing & $0$ & $-1$\\
$Y_d$ & \fund & \sing & \afund & \sing & \sing & $0$ & $1$\\
$Y_e$ & \sing & \sing & \sing & \fund & \afund & $0$ & $1$\\
\end{tabular}
\end{center}
\caption{The transformation properties of the chiral superfields and the spurions under the non-anomalous flavor symmetries preserved by the $\mu$ term. We omit discrete symmetries and a non-anomalous U$(1)_R$ which is broken by the soft terms, including the $B_{\mu}$ term.
\label{tab:flavor}}
\end{table}

The basic assumption of minimal flavor
violation~\cite{Sekhar,Lisa,MFVBuras,MFVEFT} is that the Yukawa
couplings $Y_u$, $Y_d$, and $Y_e$ are the \emph{only} spurions which
break the nonabelian SU$(3)^5$ flavor symmetry. No assumption on
baryon or lepton number is made.  Thus, while flavor non-singlet terms
may be written in the superpotential, or as soft breaking terms, their
coefficients must be built out of combinations of Yukawa couplings and
their complex conjugates in a way which respects the underlying
spurious flavor symmetry. The main new ingredient in applying MFV to SUSY theories is that the spurions also have to be assigned to representations of supersymmetry. Since the spurions $Y_{u,d,e}$ appear in the superpotential in the Yukawa terms, the most natural assumption is to assign these spurions to chiral superfields, with the expectation that in a UV completion these spurions would emerge as VEVs of some heavy chiral superfields. This assignment for the spurions ensures that  the conjugate Yukawa couplings $Y^{\dag}$ \emph{cannot}
appear in the superpotential, which will lead to a very restrictive
ansatz, both for R-parity violating terms and for higher dimensional
operators.

The MFV hypothesis can be shown to naturally suppress FCNCs~\cite{MFVEFT,MFVBuras},
thereby solving the new physics flavor problem.
It is also RGE stable, due to the spurious flavor symmetries, which prevent
flavor violating terms from being generated radiatively except those
proportional to the original spurions themselves. As explored
in~\cite{Nikolidakis:2007fc}, it is possible to impose the MFV
hypothesis on spurious (and even anomalous) U$(1)$ symmetries as
well. However, we will not do so, since the abelian symmetries are not
needed to suppress FCNCs, and furthermore, imposing such a hypothesis
will generally lead to phenomenology which is closer to the R-parity
conserving MSSM, while our primary goal is to demonstrate a viable
supersymmetric model with vastly different phenomenology.

In addition to FCNCs, low-energy CP violation (CPV) searches and
measurements also impose strong constraints on new physics.
Experimentally, CPV has been discovered only in flavor changing
processes in $K$ and $B$ decays. In the SM, this is explained by
the fact that the only source of CPV is the one physical phase of the
CKM matrix. When extending the SM, however, many new sources of CPV
can arise, both in flavor changing as well as flavor conserving
couplings. The MFV framework suppresses all new flavor-changing CPV effects, but does not address the problem
of flavor diagonal sources of CPV. In SUSY, in particular, new flavor diagonal
couplings can give rise to large EDMs, and thus the new phases cannot
be order one, and must be tuned to satisfy experimental constraints~\cite{Grossman:1997pa}. Within MFV, one solution
is to assume that all CP violating spurions come from the Yukawa matrices.
In this work, we will not consider the problem of CP violation any further, as we do not
expect that the problem will be qualitatively
different for MFV SUSY than for other MFV models~\cite{Kagan:2009bn}.

Thus, we will make the ``minimal'' assumption that the holomorphic
spurions $Y_u$, $Y_d$, $Y_e$ are the only sources of SU$(3)^5$
breaking, discarding R-parity as a means of stabilizing the
proton. This assumption, together with the holomorphy of the Yukawa
couplings, turns out to be very restrictive. It is straightforward to 
find the complete list of irreducible holomorphic flavor singlets, shown in Table~\ref{tab:inv}. The superpotential is therefore built from gauge invariant combinations of these operators. In particular, since none of these operators carry lepton number, U$(1)_L$ is an exact symmetry of the superpotential.

\begin{table}[t]
\begin{center}
\begin{tabular}{c|ccc|ccc}
& SU$(3)_C$ & SU$(2)_L$ & U$(1)_Y$ & U$(1)_B$ & U$(1)_L$ & $\mathbb{Z}_2^R$\\
\hline
$(Q Q Q)$ & \sing & $\Box\!\Box\!\Box$ & $1/2$ & $1$ & $0$ & $-$\\
$(Q Q) Q$ & $\mathbf{8}$ & $\Box$ & $1/2$ & $1$ & $0$ & $-$\\
$(Y_u \bar{u}) (Y_u \bar{u}) (Y_d \bar{d})$ & $\mathbf{8}\oplus\sing$ & \sing & $-1$ & $-1$ & $0$ & $-$\\
$(Y_u \bar{u}) (Y_d \bar{d}) (Y_d \bar{d})$ & $\mathbf{8}\oplus\sing$ & \sing & $0$ & $-1$ & $0$ & $-$\\
$\det \bar{u}$ & \sing & \sing & $-2$ & $-1$ & $0$ & $-$\\
$\det \bar{d}$ & \sing & \sing & $1$ & $-1$ & $0$ & $-$ \\
\hline
$Q Y_u \bar{u}$ & $\mathbf{8}\oplus\sing$ & \fund & $-1/2$ & $0$ & $0$ & $+$\\
$Q Y_d \bar{d}$ & $\mathbf{8}\oplus\sing$ & \fund & $1/2$ & $0$ & $0$ & $+$\\
$L Y_e \bar{e}$ & \sing & \fund & $1/2$ & $0$ & $0$ & $+$\\
\hline
$H_u$ & \sing & \fund & $1/2$ & $0$ & $0$ & $+$ \\
$H_d$ & \sing & \fund & $-1/2$ & $0$ & $0$ & $+$
\end{tabular}
\end{center}
\caption{The irreducible holomorphic flavor singlets. We omit flavor-singlet spurions (irrelevant to our analysis) as well as flavor singlets formed from $\mathrm{SU}(3)_C\times\mathrm{SU}(2)_L$ contractions of products of the operators listed here.}
\label{tab:inv}
\end{table}

While holomorphy also forbids lepton number violation in the soft
breaking $A$ terms,
lepton number violation can still occur
in the K\"{a}hler potential, and in bilinear superpotential terms,\footnote{These can be generated nonholomorphically after SUSY breaking, as shown in Appendix~\ref{app:SUSYbreaking}.} $B$ terms, and 
 the soft mass mixing term
$\tilde{L}\tilde{H_d}^{\star}+c.c.$.
However, while such terms will
play an important role when we introduce neutrino masses
in~\S\ref{sec:setupneutrinos}, in the case of massless neutrinos they
are absent for the following symmetry reason. There is a
$\mathbb{Z}_3^L \in \mathrm{SU}(3)_L\times \mathrm{SU}(3)_e$ symmetry
of the form:
\begin{equation}
L \to \omega L \;\;,\;\; \bar{e} \to \omega^{-1} \bar{e} \;\;,\;\; Y_e 
\to Y_e \; ,
\end{equation}
where $\omega \equiv e^{2 \pi i/3}$ and the other fields and spurions
are not charged under $\mathbb{Z}_3^L$. In particular,
$\mathbb{Z}_3^L$ lies within the $\mathbb{Z}_3 \times \mathbb{Z}_3$
center of $\mathrm{SU}(3)_L \times \mathrm{SU}(3)_e$, and is also a
$\mathbb{Z}_3$ subgroup of U$(1)_L$. As all spurions are neutral under
$\mathbb{Z}_3^L$, we conclude that lepton number can only be violated
in multiples of three. Soft terms of this type are not possible,
whereas the lowest-dimension $\Delta L=\pm 3$ K\"{a}hler potential
corrections are dimension eight, and are strongly suppressed for a
sufficiently high cutoff.

Since, in the absence of light unflavored fermions, proton decay requires lepton number violation, we conclude that the proton is effectively stable for massless neutrinos. Thus, proton stability will only constrain the neutrino sector, as discussed in~\S\ref{sec:consneutrinos}.\footnote{The situation changes if the gravitino (or another unflavored fermion, such as an axino) is lighter than $m_p$. We discuss the resulting constraints on $m_{3/2}$ in~\S\ref{sec:consneutrinos}.}

In addition to the R-parity conserving terms~(\ref{eqn:RPCrenormW}), MFV allows only one additional renormalizable correction to the superpotential:
\begin{equation} \label{eqn:BNVrenormW}
W_{\mathrm{BNV}} = \frac{1}{2}\, w'' (Y_u\, \bar{u}) (Y_d\, \bar{d}) (Y_d\, \bar{d}) \, ,
\end{equation}
where $w''$ is an unknown $\mathcal{O}(1)$ coefficient. In combination with the MFV structure of the soft terms, most of the interesting phenomenology of our model arises from this baryon-number and R-parity violating term.

The K\"{a}hler potential need not be canonical, and is subject to
non-universal corrections. At the renormalizable level, these take the form:
\begin{eqnarray}
K & = & Q^{\dag} \left[1+ f_Q (Y_u Y_u^{\dag}, Y_d
Y_d^{\dag})^T+h.c.\right] Q+\bar{u}^{\dag} \left[1+ Y_u^{\dag}\, f_u
(Y_u Y_u^{\dag}, Y_d Y_d^{\dag}) Y_u+h.c.\right] \bar{u}~~~~~\nonumber
\\
& & +\bar{d}^{\dag} \left[1+ Y_d^{\dag}\, f_u (Y_u Y_u^{\dag}, Y_d Y_d^{\dag}) Y_d+h.c.\right] \bar{d} \nonumber\\
& & +L^{\dag} \left[1+ f_L (Y_e Y_e^{\dag})^T+h.c.\right] L+\bar{e}^{\dag} \left[1+ f_e (Y_e^{\dag} Y_e)+h.c.\right] \bar{e} \, ,
\end{eqnarray}
where the $f_i$ are polynomials in the indicated (Hermitean)
matrices. While the renormalizable K\"{a}hler potential can be made
canonical by an appropriate change of basis, such a change of basis is
not compatible with the holomorphy of the spurions. The situation is
analogous to that of the supersymmetric beta function, where the
one-loop NSVZ result can be shown to be exact in an appropriate
holomorphic basis, but the ``physical'' all-loop beta function is
still subject to wave function renormalization, since the gauge boson
kinetic term is non-canonical in the holomorphic basis. Similarly, in
MFV SUSY the form of the superpotential is highly constrained, but the
K\"{a}hler potential is still subject to a large number of unknown
corrections. Fortunately, these unknown corrections are suppressed by
the smallness of the Yukawa couplings.

The allowed $A$ and $B$ terms are in direct correspondence with the
allowed superpotential terms, and carry the same flavor structure, except that the $A$-terms are subject to certain subleading non-holomorphic corrections:
\begin{equation}
\mathcal{L}_{\rm soft} \supset Y_u (1+Y_u^{\dag} Y_u+\ldots) \tilde{\bar{u}} (Y_d \tilde{\bar{d}}) (Y_d \tilde{\bar{d}})
+(Y_u \tilde{\bar{u}}) (Y_d \tilde{\bar{d}}) Y_d (Y_d^{\dag} Y_d+\ldots) \tilde{\bar{d}} \; ,
\end{equation}
and similar corrections to the other $A$ terms, as explained in Appendix~\ref{app:SUSYbreaking}. However, as with corrections to the K\"{a}hler potential, these corrections are suppressed by the smallness of the Yukawa couplings.

The soft breaking scalar masses have the same basic flavor structure as the K\"{a}hler terms listed above. This implies in particular that, while FCNCs can occur via squark exchange, they are suppressed by the GIM mechanism~\cite{Glashow:1970gm}, just as in the standard model. This automatic suppression of FCNCs is a universal feature of MFV scenarios. We will quantify the flavor-changing squark mass-mixings in~\S\ref{subsec:nnbarosc}.

We defer consideration of higher-dimensional operators to Appendix~\ref{app:highdimension}, where we show that such operators will give subdominant contributions to baryon-number violating processes.

\section{The baryon-number violating vertex}\label{sec:BNVvertex}

Most of the interesting phenomenology of our model arises from the interaction~(\ref{eqn:BNVrenormW}), which we now discuss in more detail.
Performing an $\mathrm{SU}(3)^5$ transformation, we choose a basis where
\begin{equation} \label{eqn:YukawaBasis}
Y_u = \frac{1}{v_u} V_{CKM}^{\dag} \begin{bmatrix} m_u & 0 & 0 \\ 0 & m_c & 0 \\ 0 & 0 & m_t \end{bmatrix} \;\;,\;\;
Y_d = \frac{1}{v_d} \begin{bmatrix} m_d & 0 & 0 \\ 0 & m_s & 0 \\ 0 & 0 & m_b \end{bmatrix} \;\;,\;\;
Y_e = \frac{1}{v_d} \begin{bmatrix} m_e & 0 & 0 \\ 0 & m_{\mu} & 0 \\ 0 & 0 & m_{\tau} \end{bmatrix} \;,
\end{equation}
where $V_{CKM}$ is the CKM matrix and $v_{u,d}=\left<H_{u,d}\right>$ are
the Higgs VEVs, with $v^2 = v_u^2 + v_d^2 \approx (174\mathrm{\
GeV})^2$ the standard model Higgs VEV. Since the Yukawa couplings
are RG dependent quantities, we should in principle evaluate them at
the squark-mass scale to estimate~(\ref{eqn:BNVrenormW}), integrate
out the superpartners, and then run the resulting couplings down to
the QCD scale. However, to obtain a rough estimate, it is sufficient
to estimate them using the following low-energy quark
masses~\cite{pdg:2010}:
\begin{eqnarray}
m_u\sim 3\mathrm{\ MeV} & \;\;,\;\; & m_c \sim 1.3\mathrm{\ GeV}
\;\;,\;\; m_t \sim 173\mathrm{\ GeV} \sim v \;\;,\nonumber \\ m_d \sim
6\mathrm{\ MeV} & \;\;,\;\; & m_s \sim 100\mathrm{\ MeV} \;\;,\;\; m_b
\sim 4\mathrm{\ GeV} \;, \label{eqn:quarkmasses}
\end{eqnarray}
together with the lepton masses:
\begin{equation}
m_e \simeq 0.511\mathrm{\ MeV} \;\;,\;\; m_{\mu} \simeq 106\mathrm{\ MeV} \;\;,\;\; m_{\tau} \simeq 1.78\mathrm{\ GeV} \;.
\end{equation}
For the magnitudes of the CKM elements, we take
\begin{equation} \label{eqn:CKMest}
V_{CKM} \sim
\begin{pmatrix}
1 & \lambda & \lambda^3/2 \\
\lambda & 1 & \lambda^2 \\
\lambda^3 & \lambda^2 & 1 \\
\end{pmatrix},
\end{equation}
where $\lambda \sim 1/5$ approximates all elements to better than $20\%$ accuracy.

The lepton and down-type Yukawa couplings depend strongly on $\tan
\beta \equiv v_u/v_d$. We consider a broad range, $3\lsim
\tan\beta\lsim 45$, where the lower bound is motivated by electroweak
symmetry breaking, and the upper bound by perturbativity of the bottom
Yukawa coupling, $y_b \lsim 1$. Consistent with the lower bound
$\tan\beta\gsim 3$, we will usually assume $\tan\beta\gg1$, which
simplifies many formulae.

Using the assumptions outlined above, we now estimate the size of the
baryon-number violating term~(\ref{eqn:BNVrenormW}), which is
conventionally written in the form:
\begin{equation}
W_{\mathrm{BNV}} = \frac{1}{2} \lambda''_{i j k} \epsilon^{abc} \bar{u}_a^i \bar{d}_b^j \bar{d}_c^k \, ,
\end{equation}
where $a,b,c$ are color indices and $i,j,k$ are the flavor indices, with summation over repeated indices understood. The factor of one-half is due to the anti-symmetry of the operator in the down-type flavor indices (which is a consequence of the color contraction). Using the basis~(\ref{eqn:YukawaBasis}), we find
\begin{equation}
\lambda''_{i j k} = w'' y^{(u)}_i y^{(d)}_j y^{(d)}_k \epsilon_{j k l} V_{i l}^{\star} \, ,
\end{equation}
where $y_i^{(u)}$ and $y_i^{(d)}$ are the up and down-type Yukawa couplings, and the coupling scales like $(\tan \beta)^2$ for large $\tan\beta$. Using the CKM estimate~(\ref{eqn:CKMest}), we find
\begin{eqnarray}
\lambda''_{usb} \sim  t^2_{\beta} \frac{m_b m_s m_u}{m_t^3} \ , \ \
& \displaystyle{\lambda''_{ubd} \sim \lambda t^2_{\beta} \frac{m_b m_d m_u}{m_t^3} }\ , \ \
& \lambda''_{uds} \sim  \lambda^3 t^2_{\beta} \frac{m_d m_s m_u}{2\, m_t^3} \ , \nonumber \\[4pt]
\lambda''_{csb} \sim \lambda t^2_{\beta} \frac{m_b m_c m_s}{m_t^3}\ , \ \
& \displaystyle{\lambda''_{cbd} \sim t^2_{\beta} \frac{m_b m_c m_d}{m_t^3}}\ , \ \
& \lambda''_{cds} \sim \lambda^2 t^2_{\beta} \frac{m_c m_d m_s}{m_t^3}\ , \nonumber \\[4pt]
\lambda''_{tsb} \sim \lambda^3 t^2_{\beta} \frac{m_b m_s}{m_t^2} \ , \ \ 
& \displaystyle{\lambda''_{tbd} \sim \lambda^2 t^2_{\beta} \frac{m_b m_d}{m_t^2}}\ , \ \
&\lambda''_{tds} \sim t^2_{\beta} \frac{m_d m_s}{m_t^2}\ . 
\end{eqnarray}
where we $t_{\beta}$ as a shorthand for $\tan\beta$. Taking the extreme value $\tan\beta = 45$, and using the quark masses~(\ref{eqn:quarkmasses}) and $\lambda\sim1/5$, we obtain the following estimates for the size of the $\lambda''_{i j k}$ coupings (for $w''=1$):
\begin{center}
\begin{tabular}{c|ccc}
& $s\, b$ & $b\, d$ & $d\, s$ \\[3pt]
\hline\\[-10pt]
$u$ & $5\times 10^{-7}$ & $6\times 10^{-9}$ & $3\times 10^{-12}$ \\[3pt]
$c$ & $4\times 10^{-5}$ & $1.2\times10^{-5}$ & $1.2\times10^{-8}$ \\[3pt]
$t$ & $2\times 10^{-4}$ & $6\times 10^{-5}$ & $4\times 10^{-5}$
\end{tabular}
\end{center}
Due to the Yukawa suppression, the largest coupling, $\lambda''_{t s
b}$, involves as many third-generation quarks as possible, without any
first generation quarks. This coupling, however, will contribute
subdominantly to low energy baryon number violation, due to the CKM
suppression required for the third generation quarks to flavor change
into first generation external state quarks. 

There are many bounds on specific combinations of RPV
couplings~\cite{Barbier:2004ez}. These bounds typically assume a generic form for the soft-masses, and thus do not necessarily apply to MFV SUSY. However, due to the flavor suppression, the
predicted values of the RPV couplings in our case are small, and all
of these bounds are satisfied.

\section{Constraints from $\Delta B = 2$ processes}\label{sec:consmassless}

The baryon number violating interaction (\ref{eqn:BNVrenormW}) will lead to baryon number violating processes which are, in theory,
observable at low energy~\cite{Zwirner}. In particular, the most stringent limits on
baryon number violation without lepton number violation come from the
lower bound on the neutron-anti-neutron oscillation
time~\cite{Abe:2011ky}
\begin{equation} \label{eq:nnbar}
\tau_{n-\bar{n}} \ge 2.44 \times 10^8 \mathrm{\ s}\, ,
\end{equation}
and from the lower bound on the partial lifetime for $p p \to K^+ K^+$ dinucleon decay~\cite{Litos:2010}
\begin{equation} \label{eq:dinuc}
\tau_{p p \to K^+ K^+} \ge 1.7\times 10^{32} \mathrm{\ yrs}\,.
\end{equation}
Both limits come from null observation of $^{16}$O decay to various final states in the Super-Kamiokande water Cherenkov detector. Present limits on other dinucleon partial lifetimes are somewhat weaker, at $\sim 10^{30} \mathrm{\ yrs}$~\cite{pdg:2010}.

In this section, we will only consider the simplest, tree-level diagrams for the processes of interest. While these will turn out the be the dominant diagrams, it is necessary to check that other contributions are subdominant. We outline a systematic scheme for doing so in Appendices~\ref{app:scheme} and \ref{app:search}.

\subsection{$n-\bar{n}$ oscillations} \label{subsec:nnbarosc}

\begin{figure}
\begin{center}
 \begin{fmffile}{nnbaroscillation}
	        \begin{fmfgraph*}(40,30)
	      \fmfstraight
	       \fmfleft{i0,i1,i2,i3,i4,i5}
	       \fmfright{o0,o1,o2,o3,o4,o5}
	        \fmf{plain}{i1,v1}
	        \fmf{plain}{i3,v3}
	        \fmf{plain}{i5,v5}
	        \fmf{plain}{v5,o5}
	        \fmf{plain,tension=0.01,label=$\tilde{g},,\tilde{N}$,label.side=left}{v1,v3}
	        \fmf{phantom}{v1,o1}
	        \fmf{phantom}{v3,o3}
	       \fmffreeze       
	       
	        \fmf{dashes,label=$\tilde{d}$}{v1,v6}
	        \fmf{dashes,label=$\tilde{b}$}{v6,v7}
	        \fmf{plain,tension=0.5}{v7,o0}
	        \fmf{plain,tension=0.5}{v7,o2}
	        \fmf{dashes,label=$\tilde{d}$}{v3,v4}
	        \fmf{dashes,label=$\tilde{b}$}{v4,v5}
	        \fmflabel{$d$}{i1}
	          \fmflabel{$d$}{i3}
	            \fmflabel{$u$}{i5}
	              \fmflabel{$\bar{d}$}{o5}
	                \fmflabel{$\bar{d}$}{o2}
	                  \fmflabel{$\bar{u}$}{o0}
	                  \fmfv{decoration.shape=cross,decoration.size=3thick}{v4}
	                   \fmfv{decoration.shape=cross,decoration.size=3thick}{v6}
	        \end{fmfgraph*}
	    \end{fmffile}
\rpar{\bar n}{21mm}{15mm}{3mm}
\lpar{n}{19mm}{16mm}{-56mm}

\end{center}
\caption{The leading contribution to $n-\bar{n}$ oscillation.\label{fig:nnbarosc}}
\end{figure}
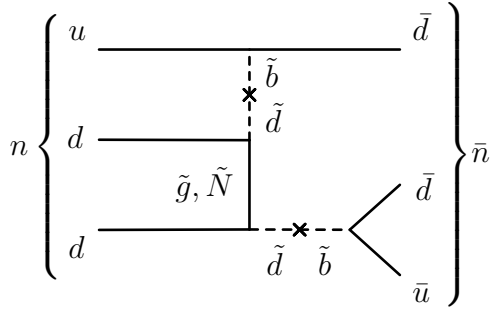

There is a unique tree-level diagram for $n-\bar{n}$ oscillations, up to crossing symmetry, the choice of the exchanged fermionic sparticle, and the squark flavors (see Fig.~\ref{fig:nnbarosc}). The down-type squarks cannot be first generation, due to the antisymmetry of $\lambda''_{i j k}$ in the last two indices. Thus, to achieve the required flavor-changing, the squarks must change flavor via mass insertions, arising from soft-terms of the form:
\begin{equation}
\mathcal{L}_{\rm soft} \supset m_{\rm soft}^2\, \tilde{Q}^{\star} \left(Y_u
Y_u^{\dag}+Y_d Y_d^{\dag}\right) \tilde{Q}+\ldots \;,
\end{equation}
where the omitted terms are higher order in the Yukawa couplings or are diagonal in the quark mass basis.

Thus, off-diagonal mass-mixing between left-handed down-type squarks of flavors $i$ and $j$ is suppressed by
\begin{equation} \label{eqn:FCNCs}
V_{i j}^{(\mathrm{neutral})} \equiv \frac{\delta m^2_{i j}}{m_{\rm soft}^2} \sim\sum_k V_{i k}^{\dag} \left[y^{(u)}_k\right]^2 V_{k j} \, ,
\end{equation}
with a similar expression for up-type squarks. The sum in~(\ref{eqn:FCNCs}) is dominated
by the third generation except in the case of $V_{u
c}^{(\mathrm{neutral})}$, where there is a competitive (though not
dominant) contribution from the second generation. We find:
\begin{eqnarray} \label{eqn:explicitFCNCs}
V_{d s}^{(\mathrm{neutral})} \sim \lambda^5 & , & V_{d b}^{(\mathrm{neutral})} \sim \lambda^3 \;\;,\;\; V_{s b}^{(\mathrm{neutral})} \sim \lambda^2 \, ,\nonumber \\
V_{u c}^{(\mathrm{neutral})} \sim y_b^2\, \lambda^5/2 & , & V_{u t}^{(\mathrm{neutral})} \sim y_b^2\, \lambda^3/2 \;\;,\;\; V_{c t}^{(\mathrm{neutral})} \sim y_b^2\, \lambda^2 \, .
\end{eqnarray}

Since the squarks in Fig.~\ref{fig:nnbarosc} are initially
right-handed, the required flavor changing is suppressed by an
additional Yukawa coupling. Depending on the initial flavor of the squark, we obtain
\begin{equation}
\tilde{b}_R \to \tilde{d}_L \; \sim \; y_b \lambda^3 \quad,\quad
\tilde{s}_R \to \tilde{d}_L \; \sim \; y_s \lambda^5 \, .
\end{equation}
As the vertex factor is also larger for a $\tilde{b}$ squark, $\tilde{b}_R \to \tilde{d}_L$ is clearly dominant.

Gathering all factors, we obtain the amplitude
\begin{equation} \label{eqn:nnbarAmp}
\mathcal{M}_{n-\bar{n}} \sim \tilde{\Lambda}\, t_{\beta}^6\, \lambda^8\ \frac{m_u^2 m_d^2 m_b^4}{m_t^8} \left( \frac{\tilde{\Lambda}}{m_{\tilde{q}}} \right)^4 \left[ g_s^2 \left( \frac{\tilde{\Lambda}}{m_{\tilde{g}}} \right)+\ldots \right],
\end{equation}
where we write the hadronic matrix element as $\tilde{\Lambda}^6$,
with $\tilde{\Lambda} \sim \Lambda_{QCD}$ in rough agreement with
the estimates of~\cite{Barbier:2004ez, Goity:1994dq}.  The omitted
terms come from neutralino, rather than gluino, exchange and can be
important if the gluino is very heavy.
 
The $n-\bar{n}$ oscillation time is approximately $t_{\rm osc} \sim \mathcal{M}^{-1}$. Therefore, assuming that the tree-level amplitude~(\ref{eqn:nnbarAmp}) gives the dominant contribution, we find
\begin{equation}
t_{\rm osc} \sim (9\times 10^9\mathrm{\ s}) \left(\frac{250\mathrm{\ MeV}}{\tilde{\Lambda}}\right)^6 \left(\frac{m_{\tilde{q}}}{100\mathrm{\ GeV}}\right)^4 \left(\frac{m_{\tilde{g}}}{100\mathrm{\ GeV}}\right)
\left(\frac{45}{\tan\beta}\right)^6 \, ,
\end{equation}
where we take $\alpha_s \equiv g_s^2/ 4 \pi \sim 0.12$. This must be
compared to the experimental bound (\ref{eq:nnbar}),
$\tau \ge 2.44 \times 10^8 \mathrm{\ s}$.  Thus, unless we have
substantially underestimated the hadronic matrix element, $n-\bar{n}$
oscillations place no constraint on our model.

\subsection{Dinucleon decay} \label{subsec:dinucleon}

\begin{figure}
\begin{center}
 \begin{fmffile}{dinucleonneutral}\hspace*{9mm}
	        \begin{fmfgraph*}(40,50)
	     \fmfstraight
	        \fmfleft{i1,i2,i3,i4,i5,i6}
	        \fmfright{o1,o2,o3,o4,o5,o6}
	        
	        \fmf{plain}{i1,o1}
	        \fmf{plain}{i2,v1}
	        \fmf{plain}{v1,o2}
	           \fmf{plain}{i5,v6}
	        \fmf{plain}{v6,o5}
	        \fmf{plain}{i6,o6}
	       	        \fmf{plain}{i3,v3}
	        \fmf{plain,tension=0.01,label=$\tilde{g},,\tilde{N}$}{v3,v4}
	        \fmf{plain}{i4,v4}
	        \fmf{phantom}{v3,o3}
	        \fmf{phantom}{v4,o4}
	        \fmffreeze
	         \fmf{dashes,tension=1,label=$\tilde{t}$}{v1,v2}
	        \fmf{dashes,tension=1,label=$\tilde{u}$}{v2,v3}
	        \fmf{dashes,tension=1,label=$\tilde{u}$}{v4,v5}
	        \fmf{dashes,tension=1,label=$\tilde{t}$}{v5,v6}
	        \fmflabel{$u$}{i1}
	        \fmflabel{$d$}{i2}
	        \fmflabel{$u$}{i3}
	        \fmflabel{$u$}{i4}
	        \fmflabel{$d$}{i5}
	        \fmflabel{$u$}{i6}
	        \fmflabel{$u$}{o1}
	        \fmflabel{$\bar{s}$}{o2}
	        \fmflabel{$\bar{s}$}{o5}
	        \fmflabel{$u$}{o6}
	        \fmfv{decoration.shape=cross,decoration.angle=0,decoration.size=3thick}{v2}
	                      \fmfv{decoration.shape=cross,decoration.angle=0,decoration.size=3thick}{v5}

	        \end{fmfgraph*}
	    \end{fmffile}  \hspace*{39mm}
 \begin{fmffile}{dinucleoncharged}
	        \begin{fmfgraph*}(40,50)
	     \fmfstraight
	        \fmfleft{i1,i2,i3,i4,i5,i6}
	        \fmfright{o1,o2,o3,o4,o5,o6}
	        
	        \fmf{plain}{i1,o1}
	        \fmf{plain}{i2,v1}
	        \fmf{plain}{v1,o2}
	           \fmf{plain}{i5,v6}
	        \fmf{plain}{v6,o5}
	        \fmf{plain}{i6,o6}
	       	        \fmf{phantom}{i3,v3}
	        \fmf{plain,tension=0.01,label=$\tilde{C}$}{v3,v4}
	        \fmf{phantom}{i4,v4}
	        \fmf{phantom}{v3,o3}
	        \fmf{phantom}{v4,o4}
	        \fmffreeze
	        \fmf{plain}{i3,v4}
	        \fmf{plain}{i4,v3}
	        \fmf{dashes,tension=1,label=$\tilde{b}$}{v1,v3}
	        \fmf{dashes,tension=1,label=$\tilde{t}$}{v4,v6}
	        \fmflabel{$u$}{i1}
	        \fmflabel{$u$}{i2}
	        \fmflabel{$d$}{i3}
	        \fmflabel{$u$}{i4}
	        \fmflabel{$d$}{i5}
	        \fmflabel{$u$}{i6}
	        \fmflabel{$u$}{o1}
	        \fmflabel{$\bar{s}$}{o2}
	        \fmflabel{$\bar{s}$}{o5}
	        \fmflabel{$u$}{o6}

	        \end{fmfgraph*}
	    \end{fmffile}
\rpar{K^+}{9mm}{45mm}{2mm}
\rpar{K^+}{9mm}{2mm}{1mm}
\rpar{K^+}{9mm}{2mm}{-87mm}
\rpar{K^+}{9mm}{45mm}{-89mm}
\lpar{p}{14mm}{7mm}{-60mm}
\lpar{p}{14mm}{7mm}{-147mm}
\lpar{p}{14mm}{40mm}{-63mm}
\lpar{p}{14mm}{40mm}{-150mm}

\end{center}
\caption{Dinucleon decay via neutral gaugino exchange (left) and
chargino exchange (right).\label{fig:dinucleondecay}}
\end{figure}
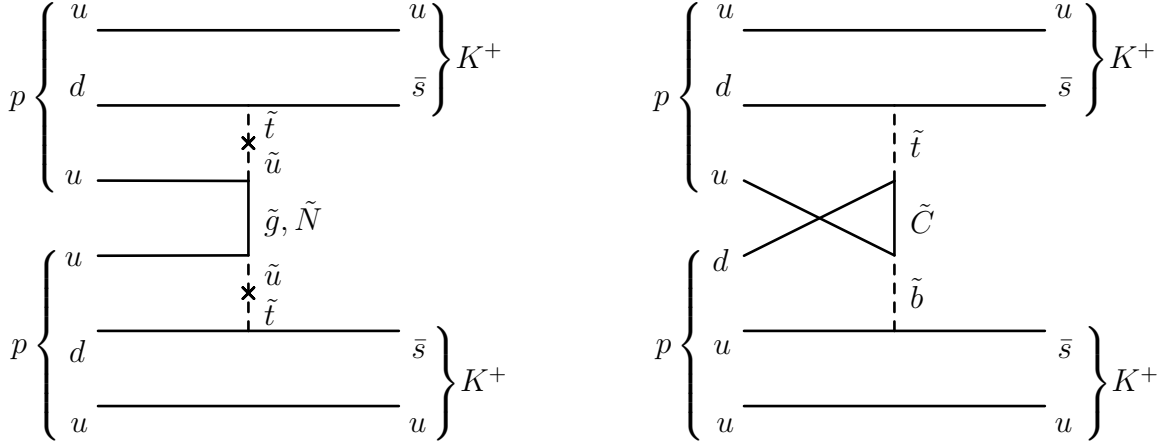

The simplest diagrams for dinucleon decay take the same form as the tree-level $n-\bar{n}$ diagram (see Fig.~\ref{fig:nnbarosc}), with the addition of two spectator quarks, as shown in Fig.~\ref{fig:dinucleondecay}. There are two possibilities, depending on whether the exchanged sparticle is a chargino or a gluino/neutralino. In the former case, the squarks undergo charged flavor changing while converting to quarks, much like quarks exchanging a $W$ boson; charge conservation then requires that one squark is up-type and the other down-type. In the latter case, the squark/quark/neutralino vertex is flavor diagonal, but neutral flavor changing via squark mass mixing is still possible.

For simplicity, we only consider diagrams of this type.\footnote{For a more systematic treatment, see Appendices~\ref{app:scheme} and~\ref{app:search}.} The external quarks must be light quarks, no more than two of which may be strange quarks. Since the quark legs do not change flavor, only $u b s$, $u b d$, $u d s$, $c d s$, and $t d s$ vertices may be used. By enumerating all possibilities, one can check that the dominant
diagram involving chargino exchange combines a $t d s$ vertex with a
$u b s$ vertex, whereas the dominant diagram involving
gluino/neutralino exchange
combines two $t d s$ vertices with $\tilde{t} \to \tilde{u}$
flavor-changing mass mixing along the squark lines. The two diagrams
are shown in Fig.~\ref{fig:dinucleondecay}, with flavor suppressions
$y_u y_d y_s^2 y_b^2 \lambda^6/2$ for the chargino exchange diagram,
and $y_d^2 y_s^2 y_b^4 \lambda^6/4$ for the gluino/neutralino exchange
diagram. Ignoring order-one factors (including gauge
couplings), the gluino/neutralino diagram is dominant if
\begin{equation}
\frac{y_d\, y_b^2}{2\, y_u} \simeq \frac{m_d}{2\, m_u} \left(\frac{m_b}{m_t}\right)^2 \tan^3 \beta \gsim 1\, .
\end{equation}
Thus, for $\tan\beta \gsim 12$ the gluino/neutralino diagram dominates; we focus on this possibility for the time being.

\begin{figure}
\begin{center}
\includegraphics[width=8cm]{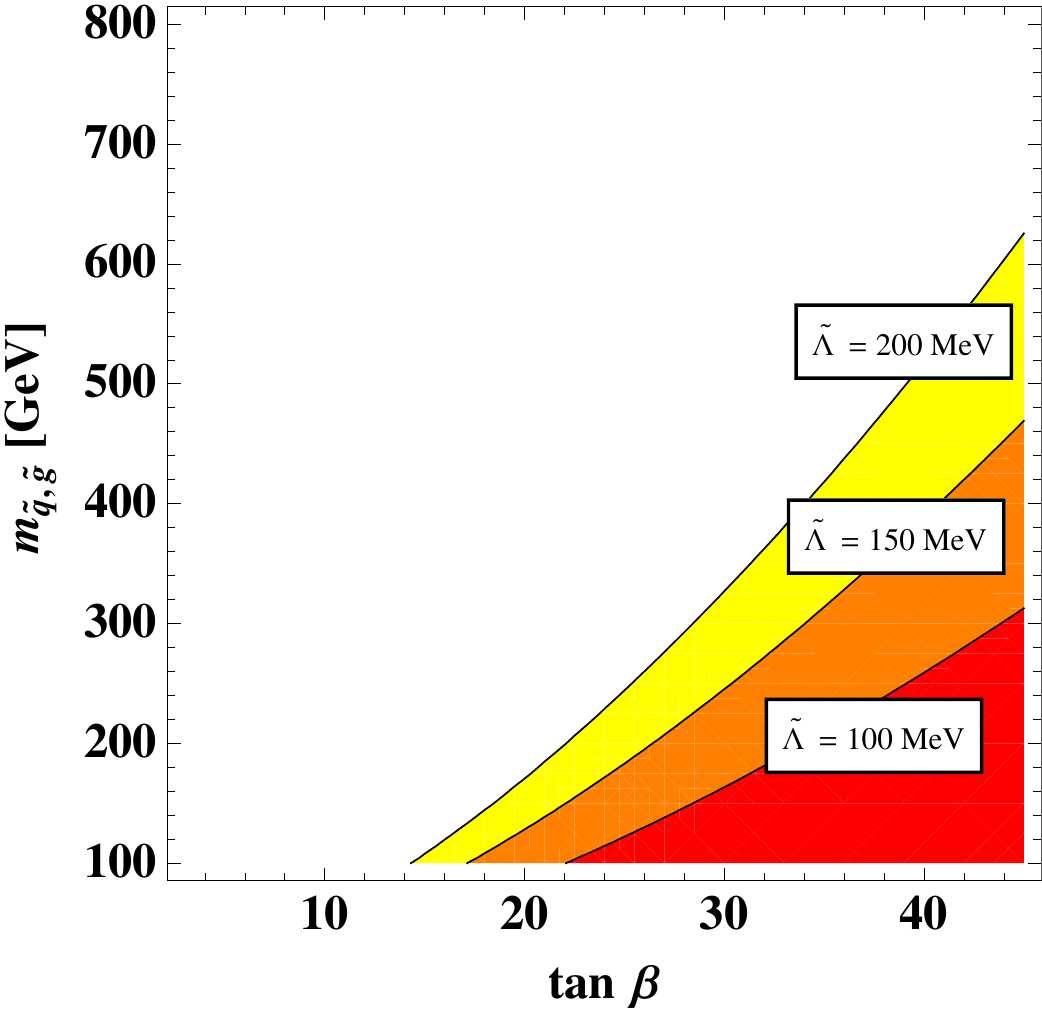}
\end{center}
\caption{Constraints on $\tan\beta$ and superparter masses due to the nonobservation of dinucleon decay. The red region is excluded assuming that $\tilde{\Lambda} \ge 100 \mathrm{\ MeV}$, whereas the orange region is also excluded when $\tilde{\Lambda} \ge 150 \mathrm{\ MeV}$, and the yellow for $\tilde{\Lambda} \ge 200 \mathrm{\ MeV}$.
\label{fig:dinucleonexclusion}}
\end{figure}

Following Goity and Sher~\cite{Goity:1994dq}, we obtain the dinucleon $N N \to K K$ width:
\begin{equation}
\Gamma \sim \rho_N \frac{128 \pi \alpha_s^2 \tilde{\Lambda}^{10}}{m_N^2 m_{\tilde{g}}^2 m_{\tilde{q}}^8} \left(\frac{\lambda^3 m_d m_s m_b^2}{2 m_t^4} \tan^4 \beta\right)^4 \, ,
\end{equation}
where $m_N \simeq m_p$ is the nucleon mass, $\rho_N \sim 0.25\mathrm{\ fm^{-3}}$ is the nucleon density, and $\tilde{\Lambda}$ is the ``hadronic scale,'' arising from the hadronic matrix element and phase-space integrals. Thus,
\begin{equation}
\tau_{N N \to K K} \sim \left(1.9\times 10^{32}\mathrm{\ yrs}\right) \left(\frac{150\mathrm{\ MeV}}{\tilde{\Lambda}}\right)^{10} \left(\frac{m_{\tilde{q},\tilde{g}}}{100\mathrm{\ GeV}}\right)^{10} \left(\frac{17}{\tan\beta}\right)^{16} \, ,
\end{equation}
where, as before, we take $\alpha_s \sim 0.12$. Comparing with the
experimental bound (\ref{eq:dinuc}), $\tau\ge 1.7\times
10^{32}\mathrm{\ yrs}$, we obtain an upper bound
\begin{equation} \label{eqn:tanbetaBound}
\tan \beta \lsim 17 \left(\frac{150\mathrm{\ MeV}}{\tilde{\Lambda}}\right)^{5/8} \left(\frac{m_{\tilde{q},\tilde{g}}}{100\mathrm{\ GeV}}\right)^{5/8} \, .
\end{equation}
This bound is illustrated in Fig.~\ref{fig:dinucleonexclusion}.

There remains considerable uncertainty in the hadronic matrix element. Goity and Sher consider values for
$\tilde{\Lambda}/m_{\tilde{q},\tilde{g}}$ between $10^{-3}$ and
$10^{-6}$~\cite{Goity:1994dq}. An earlier paper by Barbieri and
Masiero, while taking a substantially different approach, obtains a
result consistent with $\tilde{\Lambda} \sim 150\mathrm{\
MeV}$~\cite{Barbieri:1985ty}.  We will take $\tilde{\Lambda} =
150\mathrm{\ MeV}$ as a representative value. While this is somewhat
smaller than the ``natural'' $\sim \Lambda_{\rm QCD}$ scale that one
might expect, the matrix element is expected to be suppressed by
hard-core repulsion between the nucleons, motivating the yet-smaller
scales considered by~\cite{Goity:1994dq}. Due to the uncertainty
in~$\tilde{\Lambda}$, we leave the dependence on it explicit
in~(\ref{eqn:tanbetaBound}); Fig.~\ref{fig:dinucleonexclusion}
illustrates the effect of varying~$\tilde{\Lambda}$.

Assuming $m_{\tilde{q},\tilde{g}} \gsim 100\mathrm{\ GeV}$, the charged flavor-changing diagram does not alter the above bounds, since both amplitudes increase with $\tan \beta$, whereas the neutral flavor-changing diagram is already sufficiently suppressed at $\tan \beta \sim 12$, below which charged flavor-changing becomes dominant.

\section{Incorporating neutrino masses}\label{sec:setupneutrinos}

We have seen that in the absence of neutrino masses the MFV SUSY
approach approximately conserves lepton number, leaving an exact
$\mathbb{Z}_3^L$ lepton number symmetry unbroken. To introduce
neutrino masses, we therefore require additional spurions, which will
lead to additional allowed operators in the Lagrangian~\cite{Cirigliano:2005ck,Branco:2006hz}. It is
important to fully characterize such operators as, in combination with
the baryon number violating vertex (\ref{eqn:BNVrenormW}), they can
induce proton decay.

We focus on the see-saw mechanism to generate Majorana masses for the
neutrinos. We add three right-handed sterile neutrinos, $\bar N$,
which obtain Majorana masses at a heavy scale $M_R$. Through a Yukawa
coupling $Y_N$ to the left-handed neutrinos, this gives the
left-handed neutrinos a small Majorana mass of order $Y_N^2\, v^2/M_R$
upon electroweak symmetry breaking.
Due to the additional flavored field, the nonabelian spurious symmetry
of the lepton sector is extended to
$\mathrm{SU}(3)_L\times\mathrm{SU}(3)_{e}\times\mathrm{SU}(3)_{N}$.
The superpotential required to generate neutrino masses is
\begin{equation}
W_{\rm lept}= Y_e LH_d\,\bar{e}+Y_N LH_u \bar{N} +\frac{1}{2} M_N \bar N \bar N \, ,
\end{equation}
where the elements of $M_N$ are assumed to be of order $M_R$.
Thus, there are now three spurions in the lepton sector: $Y_e$, $Y_N$ and $M_N$.
The transformation properties of the leptonic sector under the spurious symmetries are shown in
Table~\ref{tab:spuriouslepton}. As before, we do not impose the MFV
hypothesis on the (spurious) U$(1)$ symmetries.

\begin{table}
\begin{center}
\begin{tabular}{c|ccc|ccc}
& SU$(3)_L$ & SU$(3)_e$ & SU(3)$_N$ & U$(1)_{B-L}$ & U$(1)_H$ & U(1)$_N$\\
\hline
$L$ &  \afund & \sing & \sing  & $-1$ & $0$ & $0$ \\
$\bar{e}$ & \sing & \fund & \sing &  $1$ & $0$ & $0$ \\
$\bar N$  & \sing & \sing & \fund & $1$ & $0$ & $1$ \\
\hline
$Y_e$ &  \fund & \afund & \sing & $0$ & $1$ & $0$ \\
$Y_N$ & \fund & \sing & \afund & $0$ & $-1$ & $-1$ \\
$M_N$ & \sing & \sing & $\overline{\symm}$ & $-2 $ & $0$ & $-2$ \\
\end{tabular}
\end{center}
\caption{The spurious leptonic flavor symmetries of the MSSM with right-handed neutrinos. We omit discrete and anomalous symmetries.\label{tab:spuriouslepton}}
\end{table}

A subtlety arises when applying the MFV hypothesis to $M_N$, since it is dimensionful. Instead, we will expand in the dimensionless spurion:
\begin{equation}
\mu_N \equiv \frac{1}{\Lambda_R} M_N \, ,
\end{equation}
where $\Lambda_R$ is an unknown heavy scale. Perturbativity of the spurion expansion requires $M_{R} \lsim
\Lambda_R$. In addition $\Lambda_R \gg m_{\rm soft}$ is required for a valid low-energy description. Otherwise, $\Lambda_R$ is an unknown scale,
which may or may not be related to other cutoff scales in the theory. 

As shown in Appendix~\ref{app:superpotential}, the
complete list of holomorphic flavor singlets involving $Y_N$, $M_N$ or $\bar{N}$ is that given
in Table~\ref{tab:MYinvariants}, where we denote the matrix of cofactors of a matrix $Y$ as
$\tilde{Y}\equiv (\det Y)\, Y^{-1}$. From these flavor singlets, only one of
the three renormalizable lepton number violating superpotential terms of~(\ref{eq:RPVrenormW}), $\lambda L L \bar{e}$, can be constructed:
\begin{equation} \label{eqn:LNVrenormW}
W^{\rm (hol)}_{\rm LNV}= \frac{1}{2 \Lambda_R}\, w \left( L L \right) \left(
\tilde{Y}_N M_{N} \tilde{Y}_{N} \right) \left(Y_e \bar{e} \right) \, ,
\end{equation}
where $w$ is an unknown $\mathcal{O}(1)$ coefficient.

\begin{table}
\begin{center}
\renewcommand{\arraystretch}{1.2}
  \begin{tabular}{c|cc|cc}
& SU$(2)_L$ & U$(1)_Y$ & U$(1)_L$ & $\mathbb{Z}_2^R$\\
\hline
    $(L L) ( \tilde{Y}_{N} M_{N}  \tilde{Y}_{N}) ( L L )$ & \sing & $- 2$ & $4$  & $+$\\
    \hline
    $(L L)  ( \tilde{Y}_{N} M_{N}  \tilde{Y}_{N} )  ( Y_e  \bar{e} )$ & \sing & $0$ & $1$ & $-$\\
    $(L L)  \tilde{Y}_{N} M_{N}  \bar{N}$ & \sing & $- 1$ & $1$ & $-$\\
    $L (Y_{N}  \tilde{M}_{N} Y_{N})  (Y_e  \bar{e})  (Y_{N}  \bar{N})$ & \fund & $1/2$ & $-1$ & $-$\\
    \hline
    $L Y_{N}  \bar{N}$ & \fund & $- 1 / 2$ & $0$ & $+$\\
    $\bar{e} Y_e  \tilde{Y}_{N} M_{N}  \bar{N}$ &
    \sing & $1$ & $-2$ & $+$\\
    $(Y_e  \bar{e}) ( \tilde{Y}_{N} M_{N} \tilde{Y}_{N}) (Y_e  \bar{e})$ & \sing & $2$ & $-2$ & $+$\\
    $L (Y_{N}  \tilde{M}_{N} Y_{N}) L$ & \symm & $- 1$ & $2$ & $+$\\
    $M_{N}  \bar{N}  \bar{N}$ & \sing & $0$ & $-2$ & $+$\\
     \end{tabular}
     \renewcommand{\arraystretch}{1.0}
\end{center}
\caption{A complete list of holomorphic flavor singlets involving $Y_{N}$ and $M_{N}$. We indicate the lepton number of the fields only, not counting that ``carried'' by the spurion $M_{N}$.\label{tab:MYinvariants}}
\end{table}

In addition, as shown in Appendix~\ref{app:SUSYbreaking}, bilinear superpotential terms, and in particular the lepton-number violating term $L H_u$, can be generated nonholomorpically after SUSY breaking. As we saw
before in the absence of neutrino masses, a $\mathbb{Z}_3^L$ symmetry
ensures that lepton number is preserved mod 3, forbidding this term. However,
while the $\mathbb{Z}_3^L$ symmetry is not broken by $Y_N$, it is
broken by $M_{N}$, which is charged under $\mathbb{Z}_3^L$.  Therefore,
bilinear lepton-number violating terms are allowed, though they necessarily
involve at least one factor of $\mu_N \sim M_R/\Lambda_R$.

The non-holomorphic corrections to the superpotential take the form:
\begin{equation} \label{eqn:LNVbilinearsup}
W^{\rm (non-hol)}_{\rm LNV} = m_{\rm soft} [\mathcal{V}^{\dag}]^a L_a H_u \; ,
\end{equation}
where there are two potentially leading contributions to the dimensionless spurion
$\mathcal{V}$:
\begin{equation}
  \mathcal{V}^{(1)}_a = \frac{1}{\Lambda_R} \varepsilon_{a b c} \left[
  \tilde{Y}_{N}^{\dag} \right]^b_i [M_{N}^{\dag}]^{i j}
  \left[Y_{N} \right]_j^c \quad , \quad \mathcal{V}^{(2)} _a =
  \frac{1}{\Lambda_R} \varepsilon_{a b c} \left[ Y_e Y_e^{\dag}
  \right]_d^b \left[ Y_{N} M_{N}^{\dag} Y_{N} \right]^{c d} \, .
\end{equation}
$\mathcal{V}^{(2)}$ contains more spurions, but if $Y_{N} \ll 1$
then the presence of the additional $Y_e$ spurions can be easily
compensated by the omission of one $Y_N$ insertion, especially at large
$\tan\beta$.

The corresponding $B$-term can also be generated, and takes the form:
\begin{equation} \label{eqn:LNVBterm}
  \mathcal{L}_{\rm soft}  \supset  m_{\mathrm{soft}}^2 [\mathcal{V}^{\dag}]^a
  \tilde{L}_a  H_u + h.c. \, ,
\end{equation}
This will lead to a left-handed sneutrino VEV 
\begin{equation} \label{eqn:snVEV}
\left<L_a \right> \sim - v_u\, \mathcal{V}_a \, ,
\end{equation}
up to an unknown $\mathcal{O}(1)$ coefficient. Inserting this VEV into
the canonical K\"{a}hler potential $L^{\dag} L$, we obtain the
gaugino/lepton mixing
\begin{equation} \label{eqn:LNVgauginomixing}
\mathcal{L} \supset - v_u\, \lambda\, (\mathcal{V}^{\dag} L) + c.c. \, .
\end{equation}
This mixing is of approximately the same order as the lepton/higgsino mixing arising from~(\ref{eqn:LNVbilinearsup}). Lepton number violation can also appear in the K\"ahler potential,
\begin{equation} \label{eqn:LNVrenormK}
K_{\rm LNV} \sim [\mathcal{V}^{\dag}]^a L_a H_d^{\dag} +
   h.c.\,,
\end{equation}
and in the correspond soft mass term. This will lead to further gaugino/lepton mixing, but proportional to $v_d$ instead of $v_u$.



In the presence of R-parity violation it is not always simple to
define which linear combination of the four fields $L_i, H_d$ is the
Higgs, and which are leptons~\cite{Banks:1995by}. The physical effects
of R-parity violation arise from a basis independent misalignment of
the different mixings between the lepton and Higgs superfields. In our
case there are several mixing terms, and cancellations can occur. As supersymmetric
sources of bilinear lepton-number violation can be eliminated
by the field redefinition $L \to L -\mathcal{V} H_d$, these cancellations will depend
on the mechanism of supersymmetry breaking.

Indeed, some cancellation may naturally occur in gauge-mediated supersymmetry breaking
models, since, due to the flavor-blind nature of gauge interactions, 
SUSY breaking effects are flavor universal,
up to RGE running and subleading corrections induced by the supersymmetric
sources of flavor-breaking.
 We do not, however, assume a particular mechanism
for SUSY breaking, and thus will take the
mixings (\ref{eqn:LNVbilinearsup}) and~(\ref{eqn:LNVgauginomixing}) to
be representative without substantial cancellation. Any such cancelation will only make
the lepton-number violating effects smaller, and so ignoring such a
possibility is a conservative assumption.

The mixing~(\ref{eqn:LNVgauginomixing}) can lead to additional
contributions to the left-handed neutrino masses via a weak-scale
see-saw mechanism. We find
\begin{equation}
\delta m_{\nu} \sim \frac{\mathcal{V}^2 v_u^2}{m_{\lambda}}\,.
\end{equation}
Imposing $|\delta m_{\nu}| \lsim 1\mathrm{\ eV}$, we
obtain an upper bound
\begin{equation}
\mathcal{V} \lsim 2\times 10^{-6} \left(\frac{m_{\lambda}}{100\mathrm{\ GeV}}\right)^{1/2}
\end{equation}
Proton decay, however, will impose a much stronger bound on
$\mathcal{V}$, and consequently the weak see-saw contribution to the
left-handed neutrino masses will be negligible.

In the above discussion, we have focused on the see-saw mechanism for generating small neutrino masses. If we instead integrate out the heavy neutrinos and consider the theory below the scale $M_R$, only one combination of the $Y_N$ and $M_N$ spurions, $Y_N M_N^{-1} Y_N^T$, is relevant for neutrino mass generation. If we ignored all other spurions built from $Y_N$ and $M_N$, taking a viewpoint that is agnostic about the high-scale mechanism for neutrino mass generation, we would obtain a theory for low-energy lepton-number violation which is more restrictive than that considered above. We have also neglected the effects of RGE running below
the scale $M_R$. While such effects can be significant in detailed
numerical calculations~\cite{Grossman:2011um}, they will not
substantially alter our order of magnitude estimates.

\section{Constraints from proton decay}\label{sec:consneutrinos}

In combination with the baryon-number violating interactions studied
in \S\ref{sec:BNVvertex} and \S\ref{sec:consmassless}, the
lepton-number violating interactions enumerated in \S\ref{sec:setupneutrinos}
will lead to a finite proton lifetime. The
strongest constraint on the proton lifetime comes from the
bound~\cite{superK:2009}
\begin{equation}
\tau_{p \to \pi^0 e^+} \ge 8.2 \times 10^{33} \mathrm{\ yrs} \, .
\end{equation}
However, this bound only constrains the partial lifetime for the particular final state $\pi^0\, e^+$. For other final states, the partial lifetime bounds are weaker, often substantially~\cite{pdg:2010}.

As we show below, MFV SUSY has a strong preference for final states with positive strangeness. Such decay modes are also strongly constrained~\cite{superK:2005,pdg:2010}:
\begin{eqnarray}
\tau_{p \to e^+\, K^0} \ge 1.0\times 10^{33} \mathrm{\ yrs} & \;,\; &
\tau_{n \to e^-\, K^+} \ge 3.2\times 10^{31} \mathrm{\ yrs} \; ,\nonumber \\
\tau_{p \to \mu^+\, K^0} \ge 1.3\times 10^{33} \mathrm{\ yrs} & \;,\; &
\tau_{n \to \mu^-\, K^+} \ge 5.7\times 10^{31} \mathrm{\ yrs} \; , \nonumber \\
\tau_{p \to \nu\, K^+} \ge 2.3\times 10^{33} \mathrm{\ yrs} & \;,\; &
\tau_{n \to \nu\, K^0} \ge 1.3 \times 10^{32} \mathrm{\ yrs} \; , \label{eqn:nucleondecaybounds}
\end{eqnarray}
where we also show the (weaker) limits on bound-neutron partial
lifetimes. There are similar bounds on some three-body decays of the
form $N \to \ell + \pi + K$.

Before discussing the constraints arising from these bounds, we first estimate the size of the coefficients of the lepton-number violating operators. We use the generic parametrization of the neutrino Yukawa couplings of Casas and Ibarra~\cite{Casas:2001sr}:
\begin{equation}
Y_N^T = \frac{1}{v_u}\, {\rm diag}\left(\sqrt{M_{R1}},\sqrt{M_{R2}},\sqrt{M_{R3}}\right)\,  R\  {\rm diag}\left(\sqrt{m_{\nu 1}}, \sqrt{m_{\nu 2}},\sqrt{m_{\nu 3}}\right)\,   U^\dagger\,,
\end{equation}
where $R$ is a complex orthogonal matrix describing mixing among the
right-handed neutrinos, $U$ is the left handed neutrino mixing matrix
giving rise to atmospheric and solar neutrino oscillations, and
$M_{Ri}$ and $m_{\nu i}$ ($i=1,2,3$) are the heavy right-handed
neutrino masses and the light left-handed neutrino masses,
respectively. The mixing angles in $U$ are large and the elements of
$U$ non-hierarchical.

Since $R$ and the right-handed neutrino masses cannot be measured at low energies, we will assume a generic flavor-structure for $Y_N$. For simplicity we will assume that the right-handed neutrinos have masses of the same magnitude, and that the left-handed neutrinos also have roughly equal masses of order $0.1$ eV, with order-one neutrino mixing angles. Substantially lighter neutrino masses would imply a more hierarchical spectrum, with small Yukawa couplings $Y_N$ and consequently more suppressed lepton-number violation, whereas substantially heavier neutrino masses begin to conflict with cosmological bounds.

The neutrino Yukawa coupling is then approximately
\begin{equation}
Y_N \sim \frac{\sqrt{M_R\, m_\nu}}{v_u} \, ,
\end{equation}
where we assume that the entire $Y_N$ matrix has elements of this order. The $L L \bar{e}$ coupling is therefore
\begin{equation}
\lambda_{i j k} \sim \frac{M_R^3\, m_{\nu}^2}{\Lambda_R\, v_u^4}\, y^{(e)}_k \, ,
\end{equation}
whereas the $\mathcal{V}$ spurions are
\begin{equation}
\mathcal{V}^{(1)}_i \sim \frac{M_R^{\frac{5}{2}} m_{\nu}^{\frac{3}{2}}}{\Lambda_R\, v_u^3} \;\; , \quad
\mathcal{V}^{(2)}_{e,\, \mu} \sim \frac{M_R^2\, m_{\nu}}{\Lambda_R\, v_u^2}\, y_{\tau}^2 \;\; , \quad \mathcal{V}^{(2)}_{\tau} \sim  \frac{M_R^2\, m_{\nu}}{\Lambda_R\, v_u^2}\, y_{\mu}^2 \, .
\end{equation}
Note that
\begin{equation}
\lambda_{i j k} \sim y^{(e)}_k Y_N\, \mathcal{V}^{(1)} \, ,
\end{equation}
up to flavor structure. Therefore, due to the smallness of the Yukawa couplings, the $L L \bar{e}$ superpotential term will be a subdominant source of lepton-number violation.

\begin{figure}
\begin{center}
\setlength{\unitlength}{0.7mm}
 \begin{fmffile}{neutrondecay}\hspace*{3mm}
	        \begin{fmfgraph*}(70,80)
	     \fmfstraight
	        \fmfleft{i0,i1,i1_5,i2,i3}
	        \fmfright{o0,o1,o1_5,o2,o3}
	        \fmf{plain}{o3,i3}
	        \fmf{fermion}{v2,i2}
	        \fmf{fermion}{v2,o2}
	        \fmf{fermion}{i1,v1}
	        \fmf{phantom}{v1,o1}
	        \fmffreeze
	        \fmf{scalar,label=$\tilde{t}_L$,label.side=left}{v1,v4}
	        \fmf{scalar,label=$\tilde{t}_R$,label.side=right}{v2,v4}
	        \fmf{fermion,label=$\tilde{C}$}{v5,v1}
	        \fmf{fermion}{v5,o0}
	        \fmf{phantom}{v5,o0}
	        \fmflabel{$d_L$}{i1}
	        \fmflabel{$d_R$}{i2}
	        \fmflabel{$u$}{i3}
	        \fmflabel{$u$}{o3}
	        \fmflabel{$\bar{s}_R$}{o2}
	        \fmflabel{$e^-_L,\mu^-_L$}{o0}
	        \fmfv{decoration.shape=cross,decoration.angle=0,decoration.size=5thick}{v4}
	             \fmfv{decoration.shape=cross,decoration.angle=-30,decoration.size=5thick}{v5}
 \end{fmfgraph*}
\end{fmffile}
\hspace*{30mm}
\begin{fmffile}{protondecay}
	        \begin{fmfgraph*}(70,80)
	     \fmfstraight
	        \fmfleft{i0,i1,i1_5,i2,i3}
	        \fmfright{o0,o1,o1_5,o2,o3}
	        \fmf{plain}{o3,i3}
	        \fmf{fermion}{v2,i2}
	        \fmf{fermion}{v2,o2}
	        \fmf{fermion}{i1,v1}
	        \fmf{phantom}{v1,o1}
	        \fmffreeze
	        \fmf{scalar,label=$\tilde{u}_L$,label.side=left}{v1,v4}
	        \fmf{scalar,label=$\tilde{t}_L$,label.side=left}{v4,v6}
	        \fmf{scalar,label=$\tilde{t}_R$,label.side=right}{v2,v6}
	        \fmf{fermion}{v5,v1}
	        \fmf{fermion}{v5,v7}
	        \fmf{fermion}{o0,v7}
	        \fmflabel{$u_L$}{i1}
	        \fmflabel{$d_R$}{i2}
	        \fmflabel{$u$}{i3}
	        \fmflabel{$u$}{o3}
	        \fmflabel{$\bar{s}_R$}{o2}
	        \fmflabel{$\bar{\nu}$}{o0}
	        \fmfv{decoration.shape=cross,decoration.angle=0,decoration.size=5thick}{v4}
	        \fmfv{decoration.shape=cross,decoration.angle=0,decoration.size=5thick}{v6}
	        \fmfv{decoration.shape=cross,decoration.angle=-30,decoration.size=5thick,label=$\tilde{N}$,label.angle=75,label.dist=8}{v5}
	        \fmfv{decoration.shape=cross,decoration.angle=-30,decoration.size=5thick}{v7}
\end{fmfgraph*}
\end{fmffile}
\rpar{K^+}{12mm}{49mm}{3mm}
\rpar{K^+}{12mm}{49mm}{-84mm}
\lpar{n}{27mm}{33mm}{-154mm}
\lpar{p}{27mm}{33mm}{-69mm}
\setlength{\unitlength}{1mm}
\end{center}
\caption{The leading charged (left) and neutral (right) flavor-changing diagrams for $n \to \ell^- K^+$ and $p \to K^+ \bar{\nu}$ nucleon decay, respectively. Arrows indicate chirality. The charged flavor-changing diagram has less flavor suppression, but suffers from a chiral suppression due to the right $\to$ right chargino propagator.\label{fig:nucleondecay}}
\end{figure}
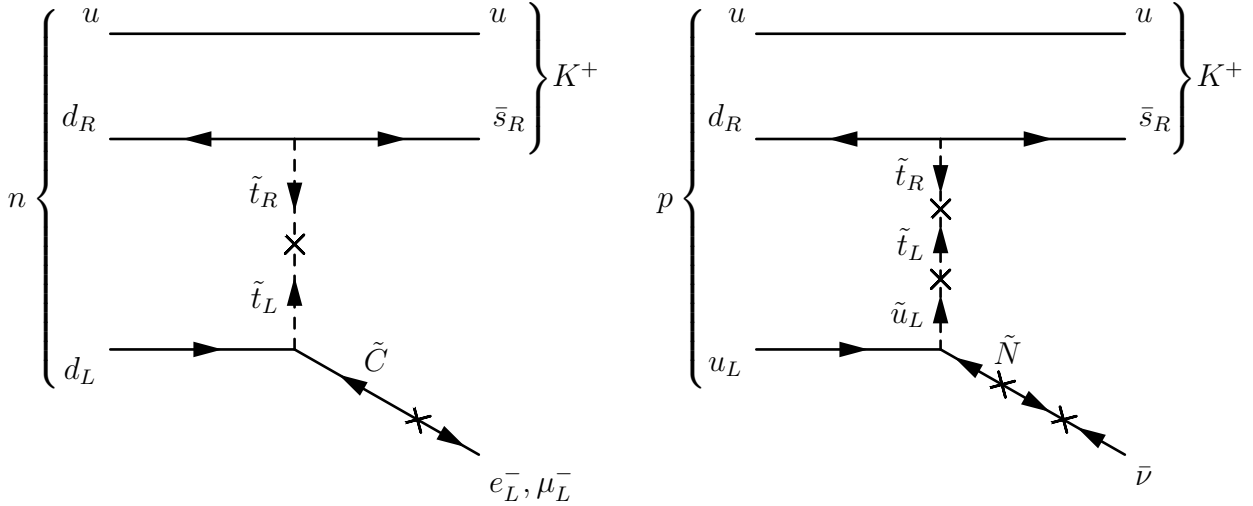

We now search for the largest possible nucleon decay diagram. The simplest diagrams for nucleon decay to a meson and a lepton are those shown in Fig.~\ref{fig:nucleondecay}, where the squark emits a chargino or neutralino, which mixes into an outgoing charged lepton or neutrino, respectively, via~(\ref{eqn:LNVgauginomixing}).\footnote{The lepton/higgsino mixing~(\ref{eqn:LNVbilinearsup}) gives another contribution to this mixing of a similar form.} Requiring the external quarks to be light, with at most one strange quark, it is straightforward to check that the leading diagram for charged lepton emission involves a $t d s$ vertex with $\tilde{t} \to d$ flavor changing at the chargino vertex, whereas the leading diagram for neutrino emission also involves a $t d s$ vertex, but with $\tilde{t} \to \tilde{u}$ mass mixing on the squark line.

The neutrino diagram has an additional flavor suppression of order
$y_b^2/2$ relative to the charged-lepton diagram. However, the latter
diagram, which leads to $n \to K^+ \mu^-$ decay, suffers from a chiral
suppression, as we illustrate in Fig.~\ref{fig:nucleondecay}. The suppression occurs because the right to right chargino propagator is roughly
$\slashed{p}/{m_{\tilde{C}}^2}$, leading to an additional suppression
of at least $\sim m_p/m_{\tilde{C}}$ relative to the right
to left propagator. This chiral suppression is not
present in the $p \to K^+
\bar{\nu}$ diagram. Combined with the stronger partial lifetime bound
for this decay mode, the latter diagram will give the strongest
constraints.

\begin{figure}
\begin{center}
\includegraphics[width=7.75cm]{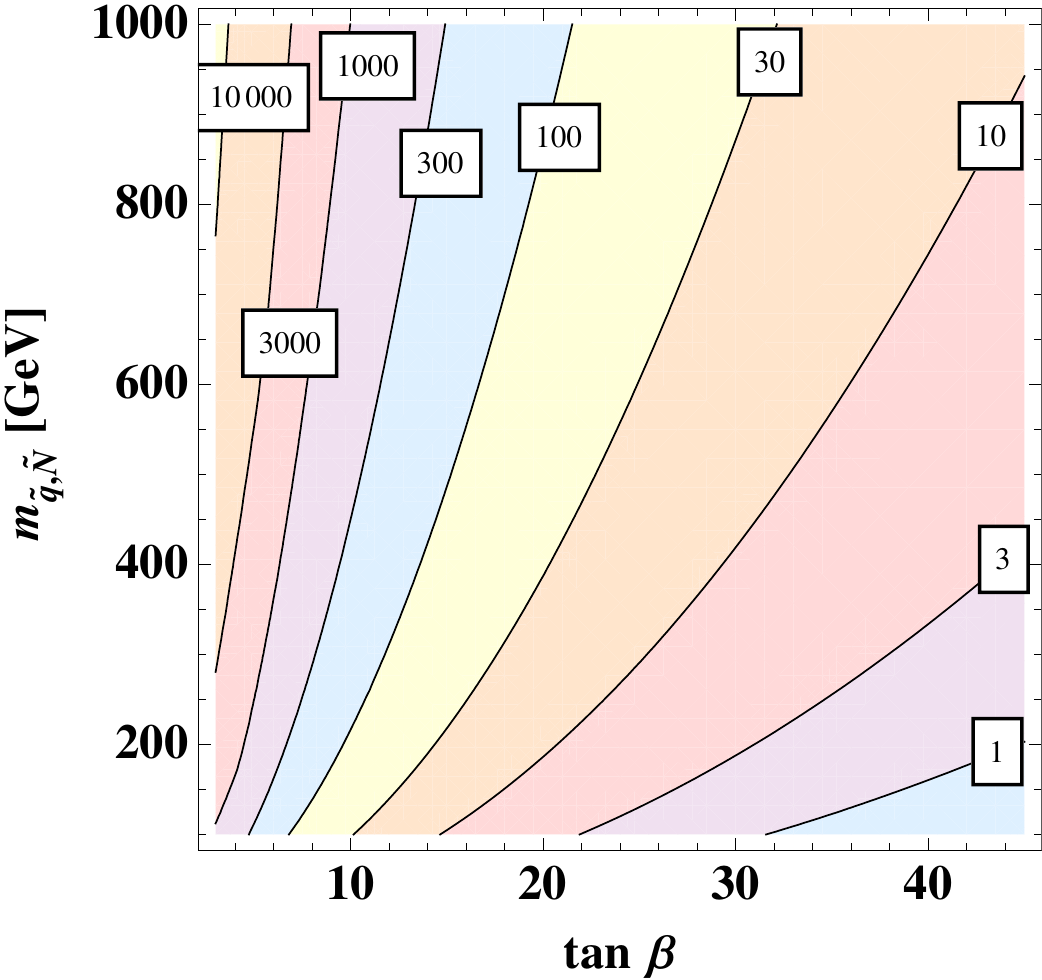}
\hspace*{0.5 cm}
\includegraphics[width=7.75cm]{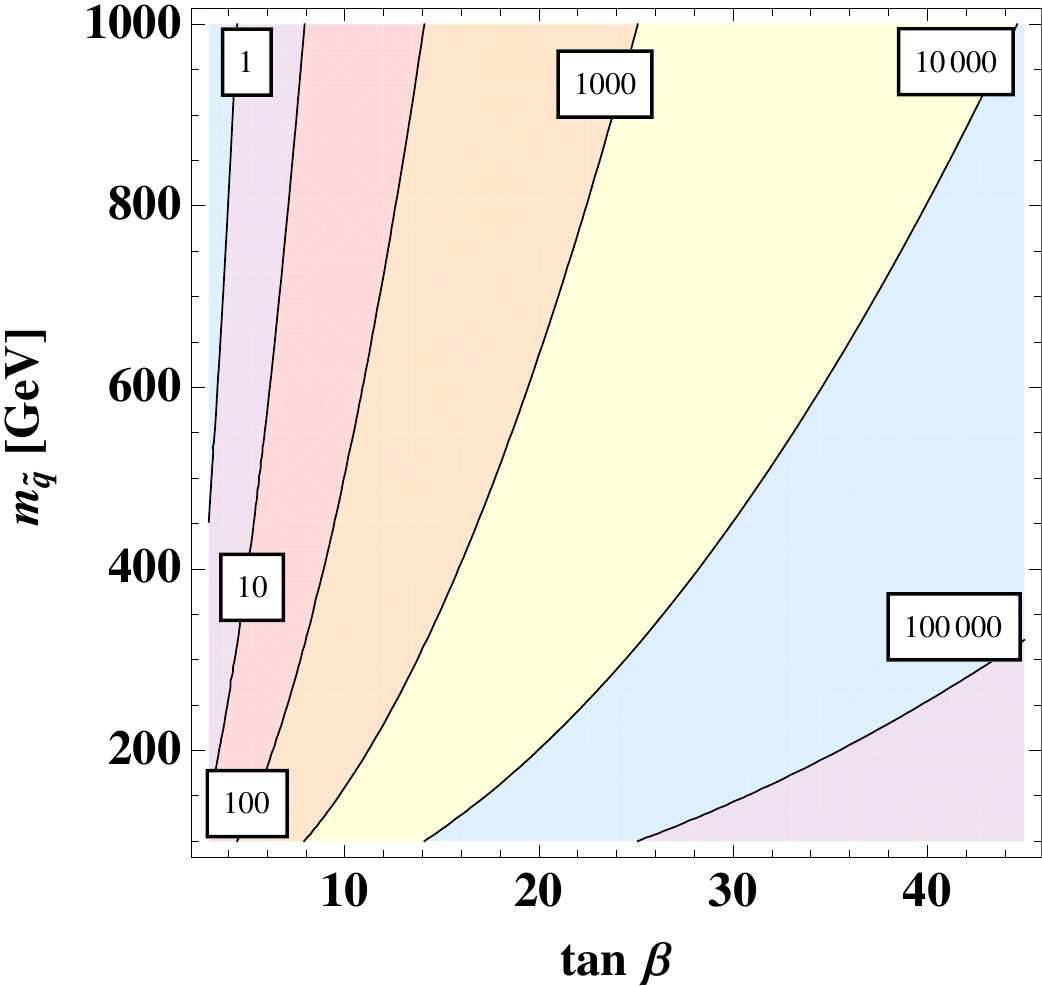}
\end{center}
\caption{Left: the upper bound on $M_R$ due to the nonobservation of nucleon decay, in units of $10^6\mathrm{\ GeV}$. For this plot, we have fixed $\Lambda_R = 10^{16}\mathrm{\ GeV}$ and $m_{\nu} = 0.1\mathrm{\ eV}$. Near the left edge, the dominant constraint comes from the $\mathcal{V}^{(1)}$ spurion; elsewhere $\mathcal{V}^{(2)}$ is dominant. Right: the approximate lower bound on $m_{3/2}$, in KeV, due to the nonobservation of $p \to K^+ \tilde{G}$.\label{fig:MRbound}}
\end{figure}

The amplitude is
\begin{equation}
\mathcal{M}_{p \to K^+ \bar{\nu}} \sim \frac{\lambda^3\, m_d\, m_s\, m_b^2}{2\, m_t^3\, m_{\tilde{N}}} \left(\frac{\tilde{\Lambda}}{m_{\tilde{q}}}\right)^2\; \mathcal{V} \tan^4 \beta \, .
\end{equation}
up to order-one mixing angles and gauge couplings, where
$\tilde{\Lambda}^2$ is a hadronic matrix element. We will take
$\tilde{\Lambda} \sim 250\mathrm{\ MeV}$, in rough agreement with
lattice computations~\cite{Cooney:2009,Aoki:2008ku}.
The width is
\begin{equation}
\Gamma \sim \frac{m_p}{8 \pi}\, |\mathcal{M}|^2 \, .
\end{equation}
Comparing with the experimental bound~(\ref{eqn:nucleondecaybounds}), we obtain
\begin{equation} \label{eqn:Vbound}
\mathcal{V}\, \tan^4 \beta \lsim (3 \times 10^{-14}) \left(\frac{m_{\tilde{q}}}{100\mathrm{\ GeV}}\right)^2 \left(\frac{m_{\tilde{N}}}{100\mathrm{\ GeV}}\right) \, .
\end{equation}
For sufficiently large $\tan\beta$, we have $\mathcal{V}^{(2)} \gg
\mathcal{V}^{(1)}$ and $\mathcal{V}^{(2)}$ gives the dominant
contribution to $\mathcal{V}$. Using $m_{\nu} = 0.1\mathrm{\ eV}$, we
then obtain the upper bound on $M_R$
\begin{equation} \label{eqn:MRbound}
M_R \lsim (3\times 10^7 \mathrm{\ GeV}) \left(\frac{10}{\tan
\beta}\right)^{3} \left(\frac{m_{\tilde{q}, \tilde{N}}}{100\mathrm{\
GeV}}\right)^{3/2} \left(\frac{\Lambda_R}{10^{16} \mathrm{\
GeV}}\right)^{1/2} \, .
\end{equation}
One can check that $\mathcal{V}^{(1)}$ gives a weaker bound than this as long as
\begin{equation}
\tan \beta \gsim 6 \left(\frac{m_{\tilde{q},\chi}}{1\mathrm{\ TeV}}\right)^{3/14}  \left(\frac{\Lambda_R}{10^{16} \mathrm{\ GeV}}\right)^{1/14} \, .
\end{equation}
Thus, for $\Lambda_R = 10^{16} \mathrm{\ GeV}$ and $m_{\tilde{q},\tilde{N}} \lsim 1\mathrm{\ TeV}$, $\mathcal{V}^{(2)}$ is dominant for $\tan \beta \gsim 6$, whereas for $\tan\beta\lsim 6$, $\mathcal{V}^{(1)}$ is dominant for sufficiently large superpartner masses. The bound on $M_R$, including both contributions, is illustrated in Fig.~\ref{fig:MRbound}.

The bound on $M_R$ depends strongly on $\Lambda_R$. For instance, if $\Lambda_R \sim 10\mathrm{\ TeV}$, the bound~(\ref{eqn:MRbound}) is reduced by six orders of magnitude. If the right-handed neutrinos are sufficiently light, they could be produced at colliders, though the Yukawa couplings are necessarily very small, so that such a scenario is unlikely to be excluded in the near future.

\begin{figure}
\begin{center}
\setlength{\unitlength}{0.5mm}
\begin{fmffile}{protondecaygravitino}
	        \begin{fmfgraph*}(70,80)
	     \fmfstraight
	        \fmfleft{i0,i1,i1_5,i2,i3}
	        \fmfright{o0,o1,o1_5,o2,o3}
	        \fmf{plain}{o3,i3}
	        \fmf{plain}{v2,i2}
	        \fmf{plain}{v2,o2}
	        \fmf{plain}{i1,v1}
	        \fmf{phantom}{v1,o1}
	        \fmffreeze
	        \fmf{dashes,label=$\tilde{u}$,label.side=left}{v1,v4}
	        \fmf{dashes,label=$\tilde{t}$,label.side=right}{v2,v4}
	        \fmf{plain}{v5,v1}
	        \fmf{plain}{v5,o0}
	        \fmf{phantom}{v5,o0}
	        \fmflabel{$u$}{i1}
	        \fmflabel{$d$}{i2}
	        \fmflabel{$u$}{i3}
	        \fmflabel{$u$}{o3}
	        \fmflabel{$\bar{s}$}{o2}
	        \fmflabel{$\tilde{G}$}{o0}
 \fmfv{decoration.shape=cross,decoration.angle=0,decoration.size=5thick}{v4}
	        
 \end{fmfgraph*}
\end{fmffile}
\rpar{K^+}{9mm}{35mm}{2mm}
\lpar{p}{20mm}{24mm}{-51mm}
\setlength{\unitlength}{0.1mm}
\end{center}
\caption{The leading contribution to $p\to K^+ \tilde{G}$ decay. \label{fig:protondecaygravitino}}
\end{figure}
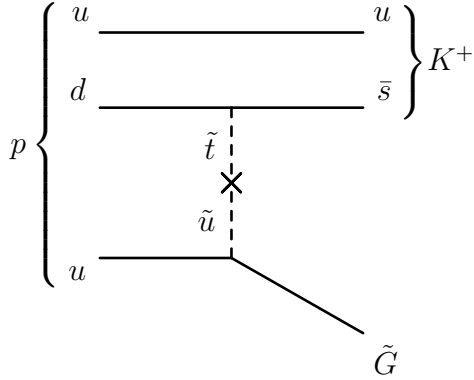

If the gravitino is sufficiently light, proton decay can proceed via the baryon-number violating vertex~(\ref{eqn:BNVrenormW}) alone, without lepton number violation~\cite{Choi:1996nk}. In particular, the gravitino is derivatively coupled to chiral superfields~\cite{Giudice:1998bp}:
\begin{equation}
\mathcal{L}_{\rm int} = - \frac{1}{\sqrt{3}\, m_{3/2}\, M_{\rm pl}}\, \bar{\psi}_L \gamma^{\mu} \gamma^{\nu} (\partial_{\mu} \tilde{G}) (D_{\nu} \phi)+c.c. \, ,
\end{equation}
where $\tilde{G}$ is the gravitino, $(\phi, \psi)$ is any chiral superfield, and $M_{\rm pl}$ is the reduced Planck mass. If kinematically allowed, the decay $p \to K^+ \tilde{G}$ will proceed via the diagram in Fig.~\ref{fig:protondecaygravitino}, with the width
\begin{equation}
\Gamma \sim \frac{m_p}{8 \pi}
\left(\frac{\tilde{\Lambda}}{m_{\tilde{q}}}\right)^4
\left(\frac{\Lambda^2}{\sqrt{3} m_{3/2} M_{\rm pl}} \right)^2 \frac{\lambda^6 m_d^2
m_s^2 m_b^4}{4 m_t^8} \tan^8 \beta \, ,
\end{equation}
where we use the same matrix element as above, replacing the momentum insertions with a characteristic energy scale, $\Lambda$.  

While we are unaware of a direct search for $p \to K^+ \tilde{G}$, for a very light gravitino $p \to K^+ \nu$ gives the same experimental signature. If we conservatively assume that the $p \to K^+ \nu$ bound~(\ref{eqn:nucleondecaybounds}) applies to $p \to K^+ \tilde{G}$ decays for any gravitino mass, we obtain an approximate lower bound on $m_{3/2}$:
\begin{equation} \label{eqn:gravitinobound}
m_{3/2} \gsim (300 \mathrm{\ KeV}) \left(\frac{300\mathrm{\ GeV}}{m_{\tilde{q}}}\right)^2 \left(\frac{\tan \beta}{10}\right)^4 \, ,
\end{equation}
where we take $\Lambda \sim \tilde{\Lambda} \sim 250\mathrm{\ MeV}$. This bound is illustrated in Fig.~\ref{fig:MRbound}.

\section{LSP decay and LHC phenomenology}\label{sec:consequences}

The phenomenology of MFV SUSY models will be very different from the
R-parity conserving MSSM, and is distinctive among R-parity violating
theories. In this section, we attempt to explore the general phenomenological features of these models. The results depend on the spectrum, and we will not attempt to exhaustively enumerate all possibilities, instead focusing on the general features for various LSPs.

We will not assume that the LSP is electrically and color
neutral; since it decays there is no particular motivation for that
requirement. Thus the LSP could be either a squark, a slepton, a
neutralino, a chargino, or the gluino. However, MFV places
restrictions on the squark and slepton masses. In particular, the mass
matrix for up-type squarks must be of the form
\begin{equation}
M_{\tilde{U}}^2 = 
\begin{pmatrix}
m_{\tilde{Q}}^2\, (1+\alpha_u Y_u Y_u^{\dag} + \alpha_d Y_d Y_d^{\dag}) + d_{u, L} & A_u\, Y_u \\
A_u^{\star}\, Y_u^{\dag}  & m_{\tilde{u}}^2\, (1+\beta_u Y_u^{\dag} Y_u)+d_{u, R}
\end{pmatrix}+\ldots \, ,
\end{equation}
where the omitted terms are higher-order in the Yukawa couplings, $A_u$ is some combination of holomorphic parameters specifying the left-right mixing (coming from the Yukawa couplings and $A$-terms), $\alpha_{u,d}$ and $\beta_u$ are non-holomorphic parameters coming from the left and right-handed squark masses, respectivley, and $d_{u, L}$ and $d_{u, R}$ are the flavor-universal $D$-term contributions to the squark masses.

Naturalness, in this context, indicates that $\alpha_{u,d}$ and $\beta_u$ should be order-one numbers, whereas $m_{\tilde{Q}}$, $m_{\tilde{u}}$, and $A_u$ are of order $m_{\rm soft}$. Thus, the leading deviations from universality will involve only the $\mathcal{O}(1)$ top Yukawa coupling, and, in particular, it is very easy to make one of the stops very light. Since other non-universal terms are suppressed by Yukawa couplings and/or CKM factors, the remaining squarks are expected to be nearly degenerate. A similar argument applies to down-type squarks, where the bottom squark can be made light. In the charged slepton sector, the leading non-universal term comes from the $y_{\tau}$ suppressed left/right mixing, implying a nearly degenerate spectrum, except at very large $\tan \beta$. The sneutrinos will be even more degenerate, since this left/right term is absent, and the leading non-universality comes from $y_{\tau}^2$ suppressed soft-mass corrections.

\begin{figure}
\vspace{0.5cm}
\begin{center}
         \begin{fmffile}{stopdecay}
	        \begin{fmfgraph*}(30,15)
	            \fmfleft{in}
	            \fmfright{o1,o2}
	            \fmflabel{ $\tilde{t}$}{in}
	            \fmflabel{ $\bar{b}$}{o1}
	            \fmflabel{ $\bar{s}$}{o2}
	            \fmf{dashes}{in,v1}
	            \fmf{plain}{v1,o1}
	            \fmf{plain}{v1,o2}
	        \end{fmfgraph*}
	    \end{fmffile}
	    \hspace*{2cm}
	    \begin{fmffile}{sbottomdecay}
	        \begin{fmfgraph*}(30,15)
	            \fmfleft{in}
	            \fmfright{o1,o2}
	            \fmflabel{ $\tilde{b}_L$}{in}
	            \fmflabel{ $\bar{t}$}{o1}
	            \fmflabel{ $\bar{s}$}{o2}
	            \fmf{dashes}{in,c1}
	            \fmf{dashes,label={ $\tilde{b}_R$}}{c1,v1}
	            \fmf{plain}{v1,o1}
	            \fmf{plain}{v1,o2}
	            \fmfv{decoration.shape=cross,decoration.size=5thick}{c1}
	        \end{fmfgraph*}
	    \end{fmffile}
  \end{center}
 \caption{The leading diagrams for stop (left) and left-handed sbottom (right) LSP decay. A right-handed sbottom decays similarly, without the mass insertion.\label{fig:stopdecay}}
 \end{figure}
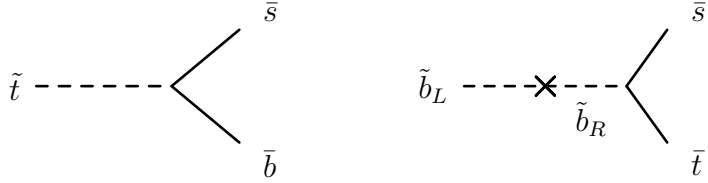

Thus, it is very natural for the stop or the sbottom to
be the LSP. A stau (or tau sneutrino) LSP, however, typically implies
a nearly degenerate spectrum, and is somewhat less natural in this
context. Other squarks or sleptons are not likely to be the LSP.

\begin{figure}
\begin{center}
\includegraphics[width=7.75cm]{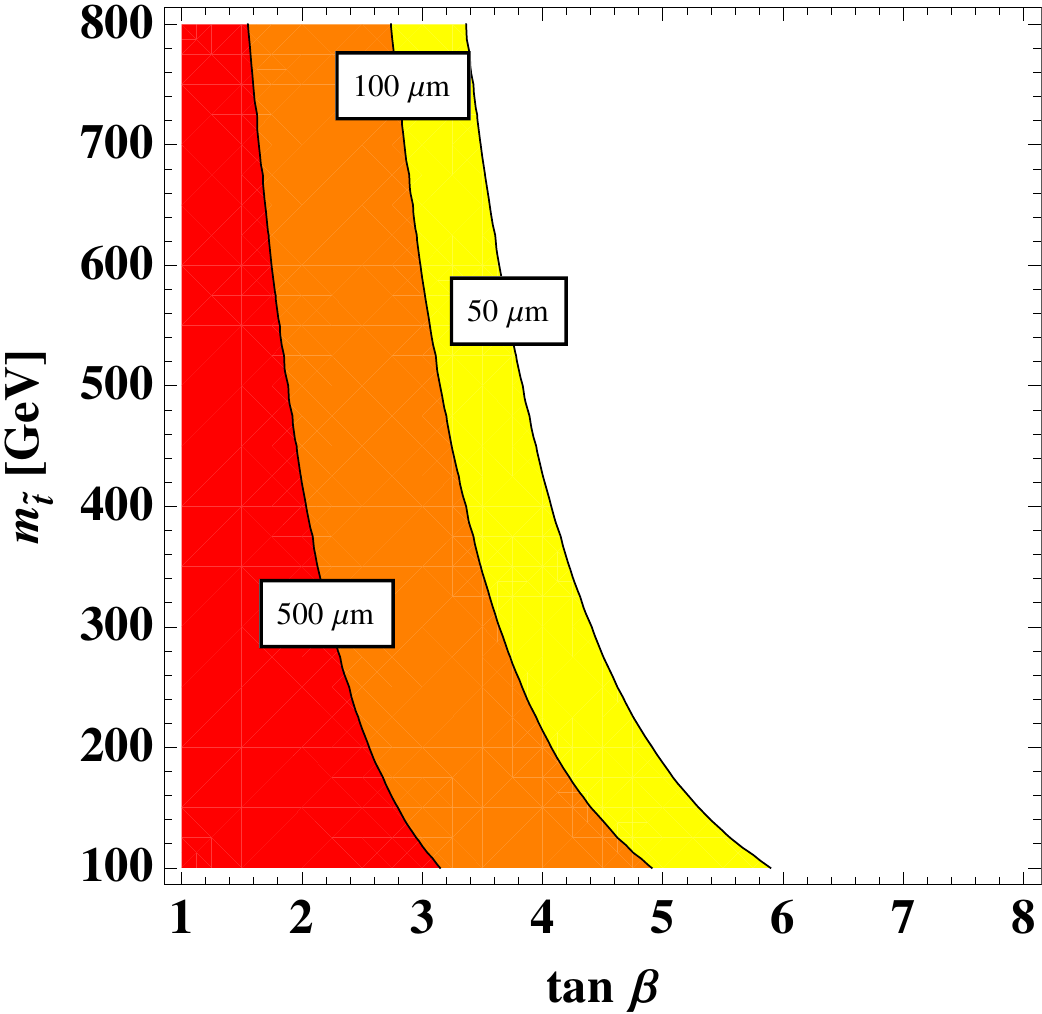}
\hspace*{0.5cm}
\includegraphics[width=7.75cm]{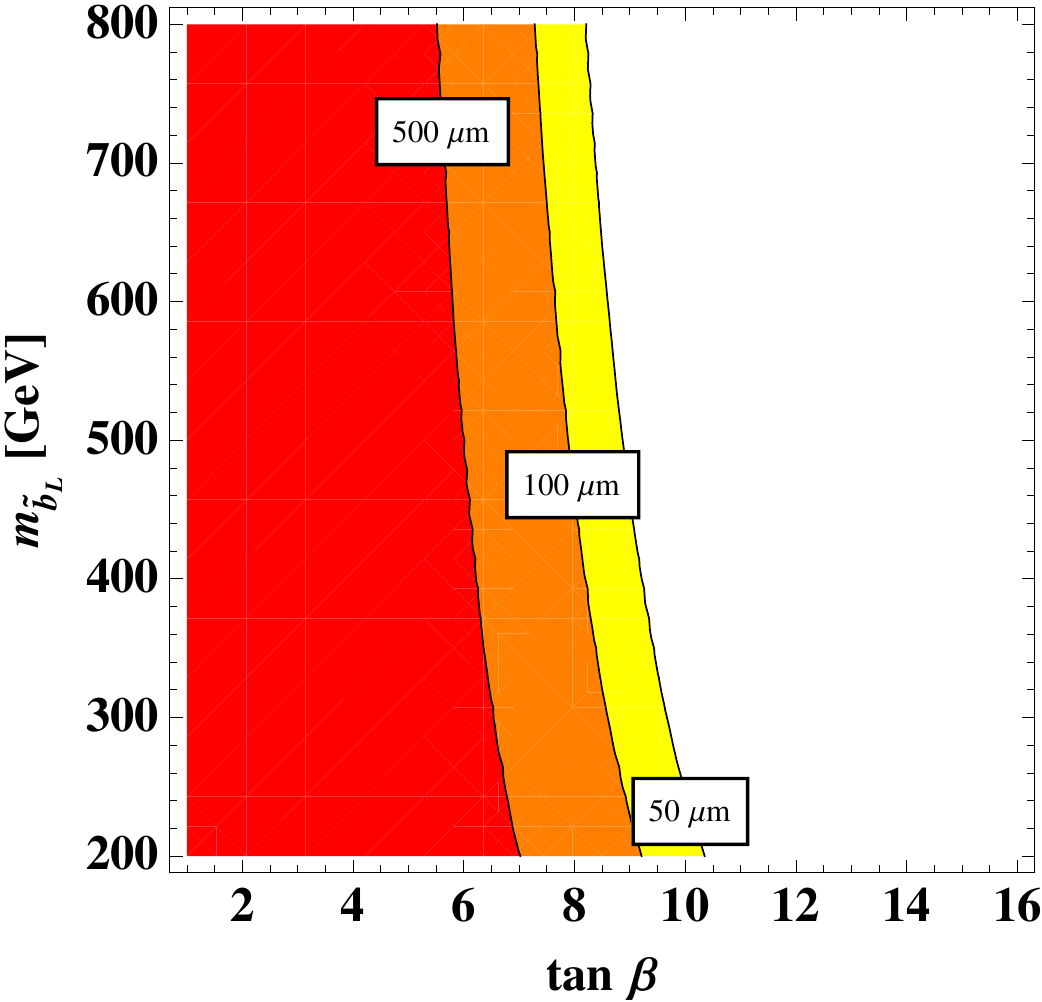}
\end{center}
\caption{The decay length ($c \tau$) of a stop (or right-handed sbottom) (left) or left-hand sbottom (right) LSP, in units of $\mu$m. Displaced vertices are expected only for small $\tan\beta$ and a light LSP.\label{fig:decaylength}}
\end{figure}

Since the largest R-parity violating operator is in the quark sector,
the most interesting scenario is when the LSP is the stop or the
sbottom. We consider the stop LSP case in detail.  The direct decay of
the stop is given by the diagram in Fig.~\ref{fig:stopdecay}.  The
partial widths $\Gamma (\tilde{t}\to \bar{d}_i\bar{d}_j)$ are given by
\begin{equation}
\Gamma_{ij}\sim \frac{m_{\tilde{t}}}{8\pi} \sin^2 \theta_{\tilde{t}} |\lambda''_{3ij}|^2 \, ,
\end{equation}
where $\theta_{\tilde{t}}$ is the stop mixing angle. To estimate the
lifetime numerically, we use the renormalized quark masses at a scale
$m_t \sim v \sim 174\mathrm{\ GeV}$, which are
approximately~\cite{Xing:2007fb, Fusaoka:1998vc}:
\begin{eqnarray}
m_u\sim 1.2\mathrm{\ MeV} & \;\;,\;\; & m_c \sim 600\mathrm{\ MeV}
\;\;,\;\; m_t \sim v \sim 174\mathrm{\ GeV} \; , \nonumber \\ m_d \sim
3\mathrm{\ MeV} & \;\;,\;\; & m_s \sim 50\mathrm{\ MeV} \;\;,\;\; m_b
\sim 2.8\mathrm{\ GeV} \;, \label{eqn:renormquarkmasses}
\end{eqnarray}
Using these masses to compute the relevant Yukawa couplings, we find a
lifetime
\begin{equation}
\tau_{\tilde{t}} \sim (2 {\rm\ \mu m}) \left( \frac{10}{\tan\beta} \right)^4 \left( \frac{300 \ {\rm GeV}}{m_{\tilde{t}}}\right) \left(\frac{1}{2\sin^2\theta_{\tilde{t}}}\right) \, .
\end{equation}
Thus no displaced vertices are expected except for very small values
of $\tan\beta$ and a very light LSP. The decay length of the stop LSP
is shown in Fig.~\ref{fig:decaylength}.

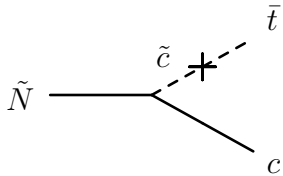
\begin{figure}
\vspace{0.5cm}
\begin{center}
\begin{fmffile}{neutralinoNLSP}
\begin{fmfgraph*}(30,15)
	\fmfleft{i1}
	\fmfright{o1,o2}
	\fmf{plain}{i1,v1}
	\fmf{plain}{v1,v2,o1}
	\fmf{dashes,label={$\tilde{c}$},l.side=left}{v1,v3}
	\fmf{dashes}{v3,o2}
	\fmfv{decoration.shape=cross,decoration.angle=45,decoration.size=5thick}{v3}
	\fmflabel{$\tilde{N}$}{i1}
	\fmflabel{$c$}{o1}
	\fmflabel{$\bar{t}$}{o2}
\end{fmfgraph*}
\end{fmffile}
\end{center}
\caption{Neutralino NLSP decay.\label{fig:neutralinoNLSPdecay}}
\end{figure}

Note that in this case one does not expect a large number of top
quarks in the final state, nor, of course, any missing energy. Roughly
90\% of decays will go to bottom and strange quarks, about 8\% to
bottom plus down, and a few percent to down plus strange.  These
branching ratios are fixed by the flavor structure. Thus, most of the
events will contain b-quarks, and a generic signal for supersymmetry
will be an overall increase in the number of events with $b$-jets, but
with possible resonances in the jet spectrum at the squark
masses. Since production of the superpartners would still be mainly
through the R-parity conserving couplings, most SUSY events would
actually end up with at least four jets, two of which are $b$-jets.
Other superpartners will first decay to the stop. For example the
neutralino is expected to decay to a stop plus charm as in
Fig.~\ref{fig:neutralinoNLSPdecay}. The neutralino lifetime for the
case of a stop LSP is given by
\begin{equation}
\Gamma_{\tilde{N}} \sim 
\frac{m_{\tilde{N}}}{8\pi} 
\,g^2 \lambda^4 \frac{m_b^4}{m_t^4} \tan^4 \beta \;\;,\;\; \tau_{\tilde{N}}\sim (10^{-19}\ {\rm s}) \left(\frac{10}{\tan \beta}\right)^4 \left(\frac{300\mathrm{\ GeV}}{m_{\tilde{N}}}\right) \, . 
\end{equation}
Thus, absent a nearly-degenerate spectrum, the other superpartners are
expected to be short-lived.

It is also possible for a bottom squark to be the LSP, decaying as shown in Fig.~\ref{fig:stopdecay}. For a right-handed sbottom, the lifetime is similar to that of a stop LSP lifetime, unless the decay is near threshold. The decay of a left-handed sbottom LSP is further suppressed by a left-right mass insertion. In this case, the partial widths $\Gamma(\tilde{b}_L \to \bar{u}_i \bar{d}_j)$ are
\begin{equation}
\Gamma_{i j} \sim \frac{m_{\tilde{b}}}{8 \pi}\, y_b^2\, |\lambda''_{i j 3}|^2 \, ,
\end{equation}
giving a total lifetime
\begin{equation}
\tau_{\tilde{b}_L} \sim (41 {\rm\ \mu m}) \left( \frac{10}{\tan\beta} \right)^6 \left( \frac{300 \ {\rm GeV}}{m_{\tilde{b}_L}}\right) \, .
\end{equation}
Thus, displaced vertices are expected at low $\tan \beta$, as illustrated in Fig.~\ref{fig:decaylength}. The phenomenology is distinct from that of a stop LSP: roughly 99\% of decays will be to top and strange or top and down quarks, with less than one percent going to charm and strange quarks, and a small fraction to other final states. Thus, an increase in top quark production is expected, with most SUSY events containing at least two top-jets. However, fewer $b$-jets will be produced, except those arising from top decays.\footnote{If $m_{\tilde{b}} \lsim m_t$, the phenomenology will be different yet again, with displaced vertices more likely due the reduced width, but no extra top production.}

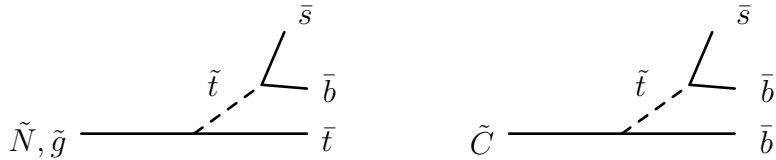
\begin{figure}
\begin{center}
\begin{fmffile}{neutralinoLSP}
\begin{fmfgraph*}(30,15)
	        \fmfbottom{i1,d1,o1}
	        \fmfright{o0,o2,o3}
	        \fmf{plain}{i1,v1,o1}
	        \fmffreeze
	        \fmf{plain}{o2,v2,o3}
	        \fmf{dashes,label={$\tilde{t}$},label.side=left}{v1,v2}
	            \fmflabel{$\tilde{N}, \tilde{g}$}{i1}
	            \fmflabel{$\bar{t}$}{o1}
	             \fmflabel{$\bar{b}$}{o2}
	            \fmflabel{$\bar{s}$}{o3}
\end{fmfgraph*}
\end{fmffile}
\hspace*{2cm}
\begin{fmffile}{charginoLSP}
\begin{fmfgraph*}(30,15)
	        \fmfbottom{i1,d1,o1}
	        \fmfright{o0,o2,o3}
	        \fmf{plain}{i1,v1,o1}
	        \fmffreeze
	        \fmf{plain}{o2,v2,o3}
	        \fmf{dashes,label={$\tilde{t}$},label.side=left}{v1,v2}
	            \fmflabel{ $\tilde{C}$}{i1}
	            \fmflabel{ $\bar{b}$}{o1}
	             \fmflabel{ $\bar{b}$}{o2}
	            \fmflabel{ $\bar{s}$}{o3}
\end{fmfgraph*}
\end{fmffile}
\end{center}
\caption{Neutralino/gluino (left) and chargino (right) LSP decays.\label{fig:neutralinoLSPdecay} }
\end{figure}

\begin{figure}
\vspace{0.3cm}
\begin{center}
\includegraphics[width=7.75cm]{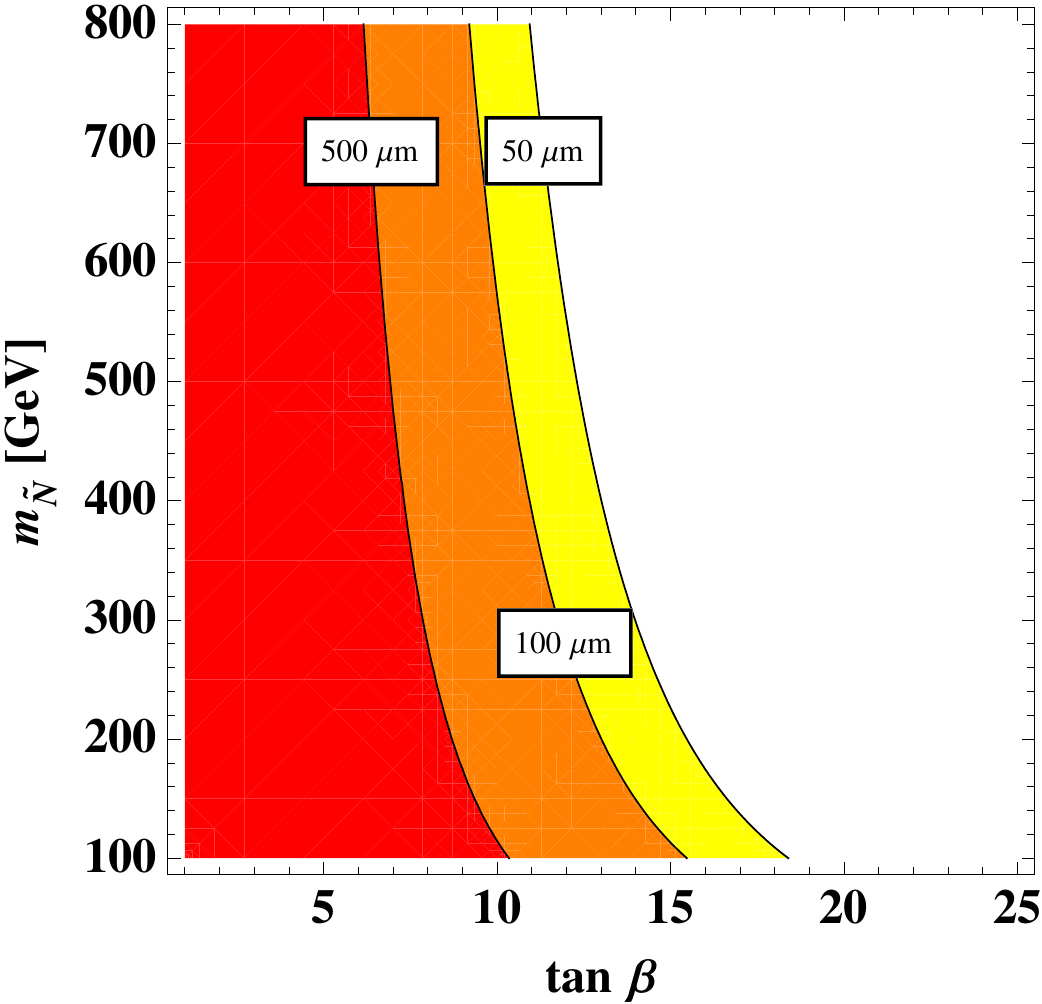}
\hspace*{0.5cm}
\includegraphics[width=7.75cm]{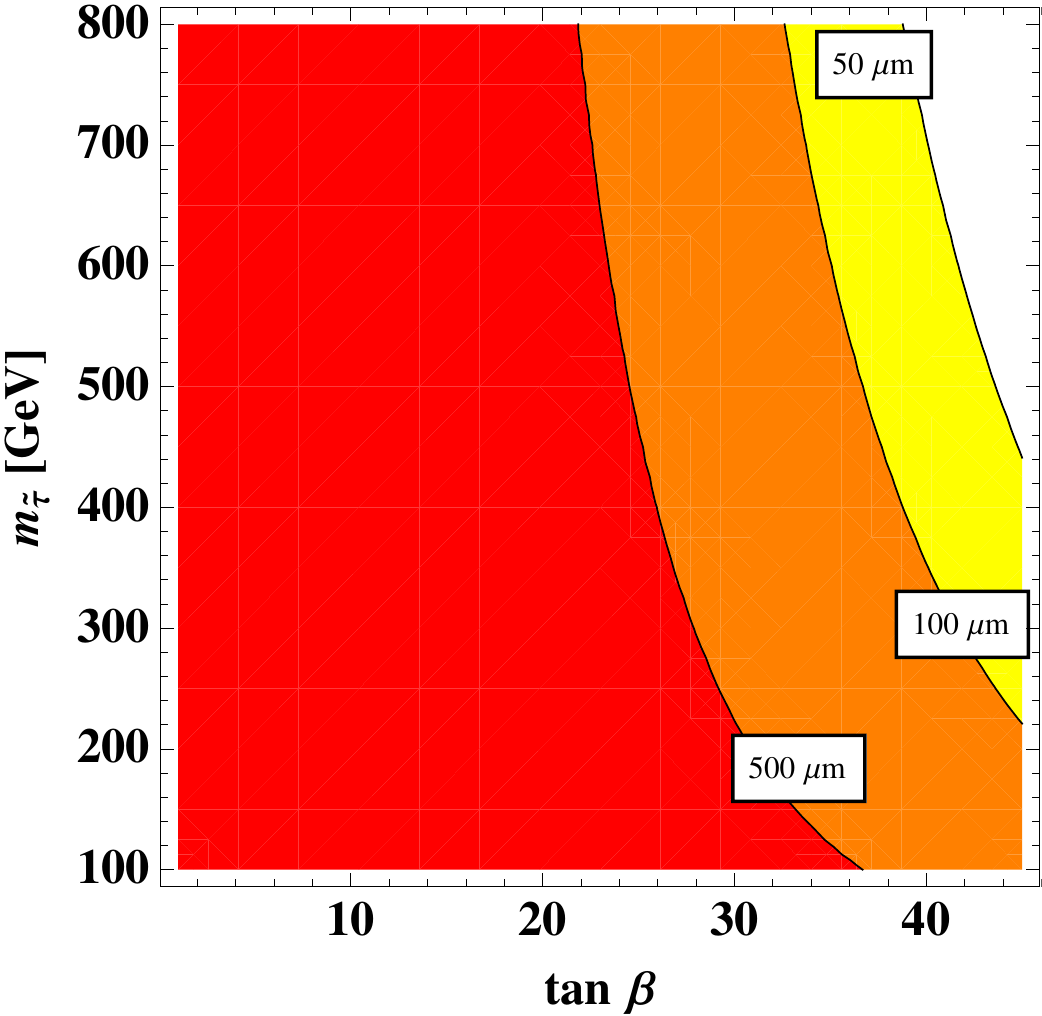}
\end{center}
\caption{The decay length ($c \tau$) of a neutralino (left) or stau (right) LSP, in units of $\mu$m. For a neutralino LSP, displaced vertices can arise in a substantial region of parameter space, whereas for the stau, they are expected nearly everywhere.\label{fig:decaylength2}}
\end{figure}

Otherwise, the LSP can be a chargino, a neutralino, or a slepton. Each
of these will give a distinct phenomenology. Assuming that the LSP is
a neutralino, its decay will be dominated by the diagram in
Fig.~\ref{fig:neutralinoLSPdecay}. The width is approximately
\begin{equation}
\Gamma_{\tilde{N}} \sim \frac{m_{\tilde{N}}}{128\, \pi^3}\, |\lambda''_{tsb}|^2 \, ,
\end{equation}
where we estimate a phase-space suppression of $1/16\pi^2$ for each additional final state particle. The lifetime is then
\begin{equation}
\tau_{\tilde{N}} \sim (12  {\rm\ \mu m}) \left( \frac{20}{\tan\beta} \right)^4 \left( \frac{300 \ {\rm GeV}}{m_{\tilde{N}}}\right) \, .
\end{equation}
As shown in Fig.~\ref{fig:decaylength2}, this scenario is much more likely to produce displaced vertices, although they can still be avoided in a sizable region of parameter space. Thus, for the case of a neutralino LSP the expected signal of SUSY would be an increase in the top production cross section (since the LSP decay involves top quarks), including potentially same-sign tops, and possibly also displaced vertices for the lights jets. A gluino LSP would decay in a very similar fashion to a neutralino LSP, whereas a chargino LSP would have a similar lifetime, but would usually decay via two $b$-jets without a top quark, as shown in Fig.~\ref{fig:neutralinoLSPdecay}. 

The case of a chargino LSP is very similar to that of a
neutralino. The one significant difference, as can be seen from Fig.~\ref{fig:neutralinoLSPdecay}, is that in the chargino case we expect no
top in the final state, and instead expect more $b$ jets.

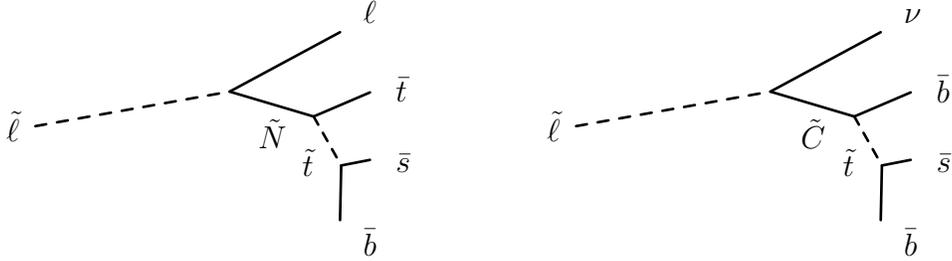
\begin{figure}
\begin{center}
 \begin{fmffile}{sleptonLSPnoMET}
	        \begin{fmfgraph*}(45,25)
	        \fmfleft{i1}
	        \fmfright{o1,o2,o3,o4}
	        \fmf{dashes}{i1,v1}
	        \fmf{plain}{v1,o4}
	        \fmf{plain,label={ $\tilde{N}$}}{v1,v2}
	        \fmf{plain}{v2,o3}
	        \fmf{dashes,label={$\tilde{t}$}}{v2,v3}
	        \fmf{plain}{v3,o2}
	        \fmf{plain}{v3,o1}
	      	   \fmflabel{ $\tilde{\ell}$}{i1}
	            \fmflabel{ $\bar{b}$}{o1}
	             \fmflabel{ $\bar{s}$}{o2}
	               \fmflabel{ $\bar{t}$}{o3}
                      \fmflabel{ $\ell$}{o4}
	        \end{fmfgraph*}
	    \end{fmffile} \hspace*{2cm}
\begin{fmffile}{sleptonLSPMET}
	        \begin{fmfgraph*}(45,25)
	        \fmfleft{i1}
	        \fmfright{o1,o2,o3,o4}
	        \fmf{dashes}{i1,v1}
	        \fmf{plain}{v1,o4}
	        \fmf{plain,label={ $\tilde{C}$}}{v1,v2}
	        \fmf{plain}{v2,o3}
	        \fmf{dashes,label={ $\tilde{t}$}}{v2,v3}
	        \fmf{plain}{v3,o2}
	        \fmf{plain}{v3,o1}
	      	   \fmflabel{ $\tilde{\ell}$}{i1}
	            \fmflabel{ $\bar{b}$}{o1}
	             \fmflabel{ $\bar{s}$}{o2}
	               \fmflabel{ $\bar{b}$}{o3}
                      \fmflabel{ $\nu$}{o4}
	        \end{fmfgraph*}
	    \end{fmffile}
   \end{center}
 \caption{Slepton LSP decay without neutrinos (left) and with neutrinos (and thus missing energy) on the right.\label{fig:sleptonLSPdecay}}
 \end{figure}

Finally, the LSP could be a slepton, mostly likely the lighter stau. This would probably be much easier to observe at the LHC. The leading decay of the stau would be a four-body decay involving top and bottom quarks, a light jet and either a lepton or missing energy, as shown in Fig.~\ref{fig:sleptonLSPdecay}. Since it is a four-body decay, the NDA estimate for the width of the stau LSP is 
\begin{equation}
\Gamma_{\tilde{\tau}} \sim \frac{m_{\tilde{\tau}}}{2048 \pi^5} |\lambda''_{tsb}|^2 \, ,
\end{equation}
with lifetime of order
\begin{equation}
\tau_{\tilde{\tau}} \sim (44  {\rm\ \mu m}) \left( \frac{45}{\tan\beta} \right)^4 \left( \frac{500 \ {\rm GeV}}{m_{\tilde{\tau}}}\right)  \, .
\end{equation}
Such long lifetimes will give displaced vertices in almost all of the relevant parameter space, as shown in Fig.~\ref{fig:decaylength2}. Thus the signal of SUSY in the case of a stau LSP would be events with displaced vertices, top and bottom quarks, and either a lepton or missing energy. 

Current searches for R-parity violating supersymmetry are not very restrictive for MFV SUSY. The more restrictive searches look for leptons among the final state particles, and set bounds on the coupling $\lambda'$: this is exactly the one vanishing in MFV SUSY. For the case of a stop LSP one could expect a resonance in the dijet searches; however the production cross section of the stop is typically about three orders of magnitudes smaller~\cite{stopproduction} than the experimental sensitivities both at the Tevatron~\cite{CDFdijet} and at the LHC~\cite{CMSdijet,ATLASdijet}. 

The more relevant searches are the ones carried out by CMS~\cite{CMSRPVsearch} (and also by CDF~\cite{CDFmultijet}): here the R-parity violating decay of the gluino in the presence of a $\bar{u}\bar{d}\bar{d}$ coupling is considered by searching for a resonance in  3-jet final states, after appropriate kinematic cuts are introduced to separate potential SUSY events from QCD background. The most stringent CMS search (using 35 pb$^{-1}$ of data) yields a bound on the gluino mass $m_{\tilde{g}}>280$ GeV. However, we should emphasize that in these models the gluino does not play an essential role. Thus even if the gluino is in the TeV energy range the model could be completely natural. While these searches are very promising, an eventual null-result of this particular experiment would not remove the motivation for these theories, since this search relies on the production of a light gluino. 

Another relevant search is for massive colored scalars in 4-jet events~\cite{Atlas4jet}. Here the four most energetic jets are paired up and a resonance in the average invariant masses of the two pairs is searched for. Stop pair production followed by decays to jets would contribute to this channel. The current bounds on the mass of a colored scalar octet using 2010 LHC data are in the $150-180$ GeV range. However, the production cross section for scalar triplets is smaller, and this bound will be substantially weakened or eliminated if applied to the stop. Better background rejection can be achieved using b-tagging, since almost all the stop quarks include at least one b-jet. A recent simulation~\cite{4jetbtag} showed that such a search at the 14 TeV LHC will be able to discover stops decaying through the $\bar{u}\bar{d}\bar{d}$ coupling up to 650 GeV with 300 fb$^{-1}$ data.  A search for a lepton together with many jets has also been suggested~\cite{Lisanti:2011tm}. This search could probe MFV SUSY if the LSP is a slepton, or if it decays to top quarks, which can produce a lepton in the final state.

Throughout this paper we have been assuming a squark mass scale of order a few-hundred GeV. This is necessary to make SUSY a natural solution of the hierarchy problem. However, in this case the Higgs mass in the simplest MSSM-type extension will usually be too light. One needs an extension of the Higgs sector, for example to NMSSM-type models, to raise the Higgs mass over the 114 GeV LEP bound. Such an extension should not significantly alter the MFV structure of the theory. For example, while the $\mathbb{Z}_3$ symmetric version of the NMSSM has restricted couplings due to the (weakly broken) discrete symmetry, the superpotential (\ref{eqn:BNVrenormW}) is $\mathbb{Z}_3$ invariant, leaving the essential features of our model intact.

One of the outstanding problems of the SM and the MSSM is the issue of baryogenesis. The Higgs mass is too high in both of these theories to account for the observed matter/antimatter asymmetry directly, and the leading explanation is baryogenesis via leptogenesis. In MFV SUSY, the appearance of the $\lambda''$ baryon number violating operator, (\ref{eqn:BNVrenormW}), opens new possibilities for baryogenesis. Several scenarios that make use of this coupling have been proposed in~\cite{Savas,Cline1,Cline2,Roulet,Sarkar}. For example the model of~\cite{Sarkar} would rely on out-of-equilibrium decays of the lightest neutralino $\tilde{N}\to\bar{u}\bar{d}\bar{d}$ and needs $\lambda''$ couplings in the $10^{-4}-10^{-3}$ range. 

\begin{figure}
\begin{center}
\setlength{\unitlength}{1.8mm}
  \begin{fmffile}{gravitinodecay}
	        \begin{fmfgraph*}(30,15)
	        \fmfleft{i1}
	        \fmfright{o1,o2}
	        \fmf{plain}{i1,v1}
	        \fmf{wiggly}{v1,v2,o1}
	        \fmf{plain,label={$\tilde{\gamma}$},label.side=left}{v1,v3}
	        \fmf{plain}{v3,o2}
	        \fmfv{decoration.shape=cross,decoration.angle=45,decoration.size=5thick}{v3}
	            \fmflabel{$\tilde{G}$}{i1}
	            \fmflabel{ $\gamma$}{o1}
	             \fmflabel{ $\nu$}{o2}
	        \end{fmfgraph*}
	    \end{fmffile}
\hspace*{2cm} \begin{fmffile}{gravitinodecayheavy}
\begin{fmfgraph*}(30,15)
	\fmfstraight
	\fmfleft{i0,i1,i2,i3,i4,i5,i6,i7,i8,i9,i10}
	\fmfright{o0,o1,o8,o2,o11,o4,o9,o10,o5,o12,o3}
	\fmf{plain}{i1,v1,o1}
	\fmffreeze
	\fmf{plain}{o2,v2,o3}
	\fmf{plain,left=0.5}{o4,o5}
	\fmf{dashes,tension=1.5,label={ $\tilde{c}$},label.side=left}{v1,v2}
	\fmflabel{ $\tilde{G}$}{i1}
	\fmflabel{ $\bar{c}$}{o1}
	\fmflabel{ $\bar{s}$}{o2}
	\fmflabel{ $\bar{b}$}{o3} 
	\fmflabel{ $\bar{u}$}{o4} 
	\fmflabel{ $u$}{o5} 
\end{fmfgraph*}
\end{fmffile}
\rpar{B^+}{8mm}{25mm}{3mm}
\rpar{\Xi_c^-}{9mm}{7mm}{2mm}
\setlength{\unitlength}{1mm}
   \end{center}
 \caption{Gravitino decay via neutrino-photino mixing (left) for gravitinos below $\sim 1$ GeV, and to hadrons (right) for masses above $\sim 1$ GeV. The illustrated hadronic decay $\tilde{G} \to B^+ \Xi_c^-$, along with other decays arising from permutations of the $c b s$ flavor labels and from changing the flavor of spectator quark, is dominant when kinematically allowed.\label{fig:gravitinodecay}}
 \end{figure}
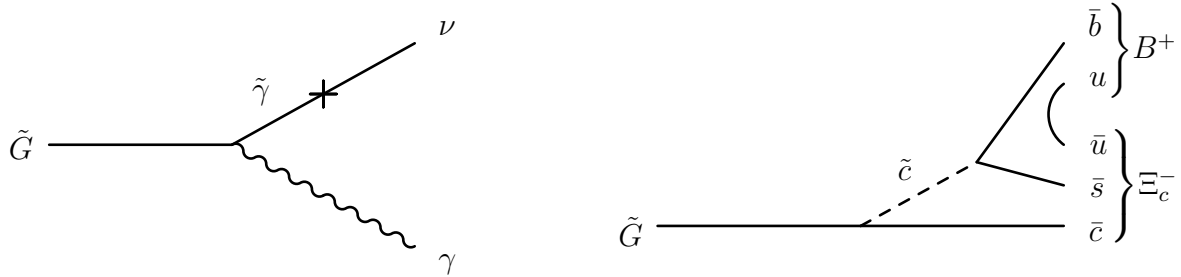

Finally we comment on dark matter. One of the main motivations for R-parity is that it provides a stable heavy superpartner, which in many cases can be a candidate for a WIMP. In MFV SUSY we are obviously forgoing this possibility. However, this does not necessarily imply that there cannot be a good dark matter candidate in these models. While we are assuming the LSP within the SM superpartners to be the stop or another sparticle, the gravitino can still be lighter and be the real LSP. A gravitino dark matter scenario within R-parity violating SUSY has been advocated in~\cite{fumi}. There it was found that the leading decay of the gravitino is $\tilde{G}\to \gamma \nu$ (see Fig.~\ref{fig:gravitinodecay}) with a width of 
\begin{equation}
\Gamma_{\tilde{G}} \sim \frac{1}{32\pi} |U_{\gamma\nu}|^2 \frac{m_{3/2}^3}{M_{Pl}^2} \, ,
\end{equation}
where $U_{\gamma\nu}$ is the photino-neutrino mixing due to the small sneutrino VEV. In our case the mixing is set by the spurion ${\mathcal V}$: $U_{\gamma\nu} \sim v_u {\mathcal V}/m_{\tilde{N}}$ where $m_{\tilde{N}}$ is a characteristic gaugino mass. Imposing the bound (\ref{eqn:Vbound}), we obtain a lower bound on the gravitino lifetime,
\begin{equation}
\tau_{\tilde{G}} \gsim (4\times 10^{39} \mathrm{\ yr}) \left( \frac{1 \ {\rm GeV}}{m_{3/2}}\right)^3 \left(\frac{300\mathrm{\ GeV}}{m_{\tilde{q}}}\right)^4 \left( \frac{\tan \beta}{10}\right)^8 \, .
\end{equation}
If the gravitino is heavier than $\sim 1$ GeV it can decay to hadrons via the R-parity violating $\bar{u}\bar{d}\bar{d}$ vertex. While the exact decay mode will depend on what is kinematically available, for $m_{3/2} \gsim 10\mathrm{\ GeV}$ all hadronic two-body decays are kinematically allowed, and the dominant mode will be that shown in Fig.~\ref{fig:gravitinodecay}. The width for the illustrated decay is
\begin{equation}
\Gamma_{\tilde{G} \to B^+ \Xi_c^-} \sim \frac{m_{3/2}^3}{24 \pi M_{\rm pl}^2} \left(\frac{\tilde{\Lambda}}{m_{\tilde{c}}}\right)^4 \frac{\lambda^2\, m_c^2 m_s^2 m_b^2}{m_t^6} \tan^4 \beta \, .
\end{equation}
Taking the matrix element to be large, $\tilde{\Lambda} \sim 1\mathrm{\ GeV}$, we find that
\begin{equation}
\tau_{\tilde{G}} \sim (2\times 10^{22} \mathrm{\ yrs}) \left(\frac{m_{\tilde{q}}}{300\mathrm{\ GeV}}\right)^4 \left(\frac{10}{\tan\beta}\right)^4 \left(\frac{100\mathrm{\ GeV}}{m_{3/2}}\right)^3 \, .
\end{equation}
In either case a gravitino LSP is generically very long lived, with a lifetime much greater than the age of the universe. Thus, the gravitino is a dark matter candidate, though more study is needed to determine if it is a realistic one.

If the gravitino is the LSP, the NLSP can either decay to jets via the R-parity violating vertex, (\ref{eqn:BNVrenormW}), or to the gravitino itself. The partial width for the simplest gravitino decay, e.g. $\tilde{t} \to t + \tilde{G}$, takes the form:
\begin{equation}
\Gamma \sim \frac{m_{\rm NLSP}^5}{24\pi m_{3/2}^2 M_{\rm pl}^2}
\end{equation}
for a squark or slepton NLSP, with a similar expression in the case of a gaugino NLSP. Thus, the rate is enhanced for a lighter gravitino, and if we assume that $m_{3/2}$ saturates the lower bound (\ref{eqn:gravitinobound}), then we obtain a branching ratio:
\begin{equation}
\frac{\Gamma_{\tilde{t} \to  t \tilde{G}}}{\Gamma_{\tilde{t} \to \mathrm{SM}}} \sim (7 \times 10^{-10})
\left(\frac{m_{\tilde{t}}}{300\mathrm{\ GeV}}\right)^8 \left(\frac{10}{\tan \beta}\right)^{12}
\end{equation}
for a stop NLSP. Thus, the branching ratio is generically small, but depends strongly on the NLSP mass and on $\tan \beta$.\footnote{For a very heavy NLSP at low $\tan \beta$, it is possible for gravitino decay to dominate, though not in a particularly promising region of parameter space.} For other NLSPs, this branching ratio is enhanced, whereas it can always be suppressed by increasing $m_{3/2}$. Depending on all the parameters, NLSP to gravitino decays could generate a significant gravitino relic density, which is of cosmological interest. We defer further consideration of this interesting topic to a future work.

\section{Conclusions}\label{sec:conclusion}\label{sec:conclusions}

We have presented an alternative approach to R-parity in supersymmetric extensions of the standard model. We have shown that imposing minimal flavor violation in a manifestly supersymmetric way  is powerful enough to reduce all baryon and lepton number violating amplitudes below current experimental bounds, while allowing a sufficiently rapid decay of the LSP such that no events with large missing transverse energy would be expected at the LHC. 

The basic MFV assumption is that the only sources of flavor violation are the SM Yukawa coupling matrices $Y_{u,d,e}$. In a supersymmetric context these spurions should be treated as VEVs of chiral superfields. The flavor symmetry together with supersymmetry will pose very stringent restrictions on the low-energy effective Lagrangian, and R-parity will be an approximate accidental symmetry. The R-parity violating terms will be determined in terms of the flavor parameters of the theory, giving an underlying theory for these parameters.~\footnote{For other theories of the R-parity violating terms see~\cite{Perez1,Perez2}.}

In the absence of neutrino masses only a single renormalizable R-parity violating flavor structure is allowed, and the proton is effectively stable, while $n-\bar{n}$ oscillations and dinucleon decay are sufficiently suppressed with mild restrictions on $\tan\beta$.
In the presence of neutrino masses there are more R-parity violating spurions, including a cubic superpotential term, and quadratic K\"ahler and soft breaking terms. Proton decay will now place a mild bound on the right handed neutrino mass scale.

The phenomenology of the model depends strongly on the nature of the LSP. The most plausible candidate for the LSP is the stop, which can  decay to two quarks via the R-parity violating superpotential term. If the LSP is a neutralino/chargino, the decay might include displaced vertices and top quarks, while a slepton LSP would most likely decay with displaced vertices, and might also involve missing energy. While the LSP is necessarily unstable in such models, a gravitino LSP is sufficiently long lived to be a dark matter candidate.

There are a number of interesting directions for future work. The constraints on MFV SUSY arising from dinucleon decay are nontrivial, and a better understanding of the relevant hadronic matrix elements would help to establish a robust set of bounds on the parameter space of the model, as well as clarifying how the model can be probed using low energy observables. Detailed collider studies are needed to determine the cleanest experimental signatures of this model at the LHC, especially in light of the various possibilities for the LSP. Furthermore, the cosmological implications for baryogenesis and dark matter should be explored in detail.

Finally, possible UV completions of the model should be explored. In R-parity conserving models, MFV is usually applied only to the SUSY breaking terms, which can be motivated by RGE evolution from flavor-universal soft terms, as in gauge mediation scenarios. In MFV SUSY, however, it is necessary to apply the MFV hypothesis to the superpotential as well, which cannot be similarly motivated. Nonetheless, an MFV structure can arise from weakly broken flavor symmetries, and constructing a well-motivated UV completion should prove to be an interesting challenge. If such a model can be found, it would give more information about the unknown flavor-singlet parameters.

\section*{Acknowledgements}

We thank Brando Bellazzini, Josh Berger, Andrey Katz, Mariangela Lisanti, Maxim Perelstein, David Shih, and Jesse Thaler for useful discussions, and Liam McAllister and Maxim Perelstein for comments on the manuscript.  We also thank Yossi Nir for pointing out Refs.~\cite{Nikolidakis:2007fc,Smith} to us. This research was supported in part by the NSF grant PHY-0757868.  C.C. and Y.G. thank the Aspen Center for Physics for its hospitality while this work was in progress. 

\appendix

\section*{Appendices --- Systematics of MFV SUSY}

MFV SUSY is a highly constrained theory, and its structure allows for a systematic approach to many problems. We outline several examples of this in these appendices. In Appendix~\ref{app:superpotential}, we show that the form of the superpotential is highly constrained by systematically classifying holomorphic flavor singlets. In Appendix~\ref{app:SUSYbreaking}, we examine the effect of supersymmetry breaking on arguments based on holomorphy. In Appendix~\ref{app:scheme}, we develop a heuristic scheme for estimating the flavor suppression of a given diagram, and in Appendix~\ref{app:search} we apply this technique to demonstrate that the diagrams presented in~\S\ref{sec:consmassless} and~\S\ref{sec:consneutrinos} are the leading contributions to low energy baryon-number violating observables. Finally, in Appendix~\ref{app:highdimension}, we show that higher dimensional baryon and lepton-number violating operators are not dangerous for a sufficiently high cutoff $\Lambda \le M_{GUT}$.

\section{Classifying holomorphic flavor singlets} \label{app:superpotential}

To classify all terms which can appear in the superpotential, we now systematically construct all holomorphic flavor singlets, treating the spurions as holomorphic. In the quark sector, the irreducible holomorphic $\mathrm{SU}(3)_u \times \mathrm{SU}(3)_d$ singlets are $Y_u \bar{u}$, $Y_d \bar{d}$, $\det \bar{u}$, $\det \bar{d}$, and the flavor-singlet spurions $\det Y_{u,d}$. Ignoring the flavor singlet spurions, and combining $Y_u \bar{u}$ and $Y_d \bar{d}$ with $Q$ to form SU$(3)_Q$ singlets, it is straightforward to show that Table~\ref{tab:inv} contains a complete list of the irreducible $\mathrm{SU}(3)_Q \times \mathrm{SU}(3)_u \times \mathrm{SU}(3)_d$ singlets.

The lepton sector is more complicated. We first write down all
possible holomorphic SU$(3)_N$ singlets. Note that for any $3\times 3$ matrix $M$
\begin{equation} \label{eqn:cofactor}
  M^{i j} M^{k l} \varepsilon_{i k m} = \varepsilon^{j l n}
  \tilde{M}_{n m} \, ,
\end{equation}
where $\tilde{M}$ is the matrix of cofactors, satisfying $\tilde{M} M = M \tilde{M} = (\det M) \mathbf{1}$.
Thus, while in general a flavor singlet can contain an arbitrary number of $\varepsilon$-tensors, by repeated application of~(\ref{eqn:cofactor}) we can reduce such a singlet to a form where no two $M_N$'s, $\tilde{M}_N$'s, $Y_N$'s, or $\tilde{Y}_N$'s are contracted with the same SU$(3)_N$ $\varepsilon$-tensor, apart from factors of $\det Y_N$ and $\det M_N$. Since at most one $\bar{N}$ can contract with a given $\varepsilon$-tensor, the only surviving $\varepsilon$-tensors must be contracted as follows:
\begin{equation}
\tilde{M}_N^{i j} \tilde{Y}_a^k \bar{N}^l \epsilon_{j k l} = -\varepsilon_{a b c} (\tilde{M}^{i j}_N Y_j^b) (Y_k^c \bar{N}^k) \, ,
\end{equation}
which is a reducible product of SU$(3)_N$ singlets. Incorporating $Y_e \bar{e}$ and $L$, we obtain a relatively short list of irreducible $\mathrm{SU}(3)_N \times \mathrm{SU}(3)_e$ singlets, as shown in Table~\ref{tab:NEsinglets}.

\begin{table}
\begin{center}
\renewcommand{\arraystretch}{1.15}
  \begin{tabular}{c|cc|cc}
    & SU$(2)_L$ & U$(1)_Y$ & SU$(3)_L$ & $\mathbb{Z}_2^R$\\
    \hline
    $\bar{N} M_N \bar{N}$ & \sing & $0$ & $\mathbf{1}$ & $+$\\
    $Y_N  \bar{N}$ & \sing & $0$ & \fund & $-$\\
    $Y_e \bar{e}$ &\sing & $1$ & \fund & $-$\\
    $L$ & \fund & $-1/2$ & \afund & $-$ \\
    $\tilde{Y}_N M_N  \bar{N}$ & \sing & $0$ & \afund & $-$\\
    $\tilde{Y}_N M_N \tilde{Y}_N$ & \sing & $0$ & $\overline{\symm}$ & $+$\\
    $Y_N \tilde{M}_N Y_N$ & \sing & $0$ & \symm & $+$
  \end{tabular}
\renewcommand{\arraystretch}{1.0}
\end{center}
\caption{The irreducible $\mathrm{SU}(3)_N \times \mathrm{SU}(3)_e$ singlets (we omit flavor-singlet spurions.)\label{tab:NEsinglets}}
\end{table}

The next step is to classify irreducible SU$(3)_L$ singlets. Note that
\begin{equation}
(\tilde{Y}_N M_N \tilde{Y}_N) (Y_N \tilde{M}_N Y_N) = (\det Y_N)^2 (\det M_N) \mathbf{1} \, .
\end{equation}
Thus, up to normalization, $\tilde{Y}_N M_N \tilde{Y}_N$ is the matrix of cofactors of $Y_N \tilde{M}_N Y_N$, and we can omit singlets containing more than one of either contracting with the same SU$(3)_L$ $\varepsilon$-tensor. There is then a finite list of possible irreducible flavor singlets. Of these, some will be reducible
due to the identities satisfied by $Y_N$ and $\tilde{Y}_N$ and $M_N$ and
$\tilde{M}_N$. For instance, any contraction involving $Y_N \tilde{Y}_N$ or
$\tilde{Y}_N Y_N$ is obviously reducible, since $Y_N \tilde{Y}_N = \tilde{Y}_N Y_N = (\det Y_N) \mathbf{1}$. Furthermore, certain $\varepsilon$-tensor
contractions of $Y_N$ with itself or $\tilde{Y}_N$ with itself will be
reducible. In particular, we have
\begin{eqnarray}
  (Y_N  \bar{N} ) ( Y_N  \tilde{M}_N Y_N
  )  ( Y_N  \bar{N} ) & \sim & 
  \bar{N} 
  \tilde{Y}_N  \tilde{M}_N  \tilde{Y}_N  \bar{N} \nonumber \\
  & \sim & 
  ( \bar{N} M_N  \tilde{Y}_N )  ( \tilde{Y}_N
  M_N  \bar{N} ) - ( \bar{N} M_N  \bar{N} ) 
  ( \tilde{Y}_N M_N  \tilde{Y}_N ) \nonumber \;\; , \\
  ( \tilde{Y}_N M_N  \tilde{Y}_N)  \tilde{Y}_N M_N 
  \bar{N} & \sim &
  ( \det Y_N )  (M_N  \tilde{Y}_N) 
  ( M_N  \bar{N} ) Y_N \; \sim \;  ( \det Y_N
  ) Y_N  \tilde{M}_N  \tilde{Y}_N  \bar{N} \nonumber \\
   & \sim &
  ( \det Y_N )  ( Y_N  \tilde{M}_N Y_N )
  ( Y_N  \bar{N} ) \;\; ,
\end{eqnarray}
up to unimportant factors.

Keeping these reductions in mind, it is straightforward to verify that Table~\ref{tab:MYinvariants} contains a complete list of $\mathrm{SU}(3)_L \times \mathrm{SU}(3)_e \times \mathrm{SU}(3)_N$ invariants, apart from $L Y_e \bar{e}$, which appears in Table~\ref{tab:inv}.

\section{Nonholomorphic operators from SUSY breaking} \label{app:SUSYbreaking}

In the absence of supersymmetry breaking, the superpotential is constrained to
be holomorphic, and only holomorphic combinations of spurions can appear
there. We now explore the role of supersymmetry breaking in introducing
nonholomorphic spurion combinations into the superpotential. To keep the
discussion of supersymmetry breaking generic, we introduce a
supersymmetry-breaking spurion $X$, a chiral superfield which acquires an
$F$-term vev $\langle X \rangle_F = F$. We assume that $X$ couples to the MSSM
fields via nonrenormalizable operators, where the cutoff $M$ is the messenger
scale.

The resulting soft supersymmetry breaking terms will appear a scale
$m_{\mathrm{soft}} \sim F / M$. In particular, since we assume the absence of
renormalizable couplings between $X$ and the MSSM, the leading contributions
to supersymmetry breaking come from the superpotential interactions
\begin{equation}
  W_{\not{\mathrm{SUSY}}} \supset \frac{X}{M} A_{i j k} \Phi^i \Phi^j \Phi^k +
  \frac{X}{M} M_{\lambda}^{\left( i \right)} \mathrm{Tr} W^2_{\left( i \right)}
\end{equation}
and the K\"ahler potential interactions:
\begin{equation}
  K_{\not{\mathrm{SUSY}}} \supset \frac{X^{\dag}}{M}  \tilde{\mu}_{i j} \Phi^i
  \Phi^j + \frac{X}{M}  \tilde{J}_i^j \Phi^i \Phi_j^{\dag} + \frac{X^{\dag}
  X}{M^2}  \tilde{B}_{i j} \Phi^i \Phi^j + c.c. + \frac{X^{\dag} X}{M^2} 
  \tilde{M}_i^j \Phi^i \Phi_j^{\dag}
\end{equation}
where nonholomorphic couplings are denoted with a tilde. The couplings $A_{i j
k}$ and $M_{\lambda}$ generate $A$-terms and gaugino masses, whereas
$\tilde{B}_{i j}$ and $\tilde{M}_i^j$ generate $B$-terms and soft-masses,
$\tilde{\mu}_{i j}$ generates bilinear superpotential terms, and
$\tilde{J}_i^j$ gives rise to a scalar/F-term mixing, the effects of which we
discuss in detail below. In singlet extensions of the MSSM, including the
NMSSM and see-saw models, supersymmetry breaking tadpoles can also arise:
\begin{equation}
  K_{\not{\mathrm{SUSY}}}^{\text{(tad)}} = \frac{X^{\dag} X}{M}  \tilde{E}_i
  \Phi^i
\end{equation}
where the dimensionful coefficient is large, $F^2 / M \sim M m_{\mathrm{soft}}^2
\gg m_{\mathrm{soft}}^3$. While these tadpoles are potentially problematic,
whether they are generated and at what level will depend on the particular
model of supersymmetry breaking. We will assume that they are suppressed by
some mechanism, and will not consider them further.{\footnote{For instance, a
right-handed snuetrino tadpole is forbidden by $\mathbb{Z}_3^{\left( L
\right)}$ in the case of Dirac neutrino masses $\left( M_N = 0 \right)$.}}

Thus, we conclude that $A$-terms are generated holomorphically, whereas the
other soft terms are generated non-holomorphically. Furthermore,
nonholomorphic bilinear couplings can appear in the superpotential at the
scale $m_{\mathrm{soft}}$. Nonholomorphic contributions to the $A$-terms and
trilinear superpotential terms are suppressed. The leading contributions arise
from the interactions
\[ K_{\not{\mathrm{SUSY}}} \supset \frac{X^{\dag}}{M^2}  \tilde{\lambda}_{i j k}
   \Phi^i \Phi^j \Phi^k + \frac{X X^{\dag}}{M^3}  \tilde{A}_{i j k} \Phi^i
   \Phi^j \Phi^k \]
which are suppressed by $\mathcal{O} \left( m_{\mathrm{soft}} / M \right)$
relative to the leading holomorphic contributions.

So far we have ignored the nonholomorphic scalar/F-term mixing
$\tilde{J}_i^j$. We will show that these couplings give rise to nongeneric
nonholomorphic contributions to the $A$-terms after a field redefinition,
similar in form to (nonholomorphic) wavefunction renormalization effects.

We first write the renormalizable superpotential and K\"ahler potential in
the form:
\begin{eqnarray*}
  W & = & m_{\mathrm{soft}} \mu_{i j} \Phi^i \Phi^j + \lambda_{i j k} \Phi^i
  \Phi^j \Phi^k\\
  K & = & \tilde{K}_i^j \Phi^i \Phi_j^{\dag}
\end{eqnarray*}
where $\tilde{K}_i^j$ is the Hermitean positive-definite K\"ahler metric. (Note that
we cannot in general set $\tilde{K}_i^j = \delta_i^j$ by a field redefinition
without introducing nonholomorphic couplings into the super\-potential.)
The
scalar/F-term mixing can be eliminated by redefining
\[ \Phi^i \rightarrow \Phi^i + \frac{X}{M}  \tilde{P}^i_j \Phi^j \]
for $\tilde{P}^i_j = - \left[ \tilde{K}^{- 1} \right]^i_k  \tilde{J}^k_j$.
This redefinition produces additional $A$-terms of the form:
\[ W_{\not{\mathrm{SUSY}}} \supset \frac{X}{M}  \left[ \lambda_{l j k} 
   \tilde{P}^l_i + \lambda_{i l k}  \tilde{P}^l_j + \lambda_{i j l} 
   \tilde{P}^l_k \right] \Phi^i \Phi^j \Phi^k \]
as well as corrections to the soft-masses and $B$-terms.

By contrast, writing
the K\"ahler potential in the form
\[ \tilde{K}^i_j = \delta^i_j + \tilde{k}^i_j \]
and assuming that $\tilde{k}^i_j$ is a subleading correction, we obtain
similar nonholomorphic corrections to the superpotential itself (as well as
the $A$-terms) upon moving to a canonical basis. Thus, we conclude that the
qualitative effects of nonholomorphic scalar/F-term mixing are captured by
nonholomorphic corrections to the K\"ahler potential, though $\tilde{J}_j^i$
leads to some additional ``splitting'' between the $A$-terms and
superpotential terms.

\section{A heuristic estimation scheme} \label{app:scheme}

In~\S\ref{sec:consmassless} and~\S\ref{sec:consneutrinos}, we estimated the dominant contribution to low-energy baryon-number violating processes by choosing the simplest diagrams and then finding the dominant flavor structure. The resulting diagrams were heavily suppressed by Yukawa couplings, CKM factors, and heavy propagators. Thus, in principle other diagrams could give competitive contributions. However, classifying all possible diagrams is a difficult task. Instead, we develop a scheme to estimate the flavor-suppression of a diagram based on its flavor structure alone. This will allow us to isolate potentially competitive diagrams, which can then be computed by more conventional means.

To do so, it is helpful to reinterpret a Feynman diagram for a candidate process in terms of the flow of ``flavor,'' i.e. of $\mathrm{SU}(3)_Q\times \mathrm{SU}(3)_u \times \mathrm{SU}(3)_d$ charge. If quarks and squarks carry ``flavor'' and anti-quarks and anti-squarks carry ``anti-flavor,'' then flavor can only be created or destroyed at baryon number violating vertices, such as~(\ref{eqn:BNVrenormW}). Otherwise, the rest of the diagram contains unbroken flavor lines, which either form closed loops or join to external quark lines.

Along flavor lines, flavor is altered through left $\leftrightarrow$ right mixing, charged CKM mixing, and neutral squark mass mixing, where each subprocess has an associated cost. In particular, for squarks, left $\leftrightarrow$ right mixing is suppressed by the associated Yukawa coupling, whereas charged-current flavor changing (on left-handed squarks) is CKM suppressed. FCNCs are suppressed by~(\ref{eqn:explicitFCNCs}), and flavor changing of right-handed squarks is suppressed by the associated Yukawa couplings to convert them to left-handed squarks, together with the suppression for left-handed flavor changing.

If we assume similar suppressions for flavor-changing processes involving quarks, we obtain a useful heuristic estimate scheme for the MFV-dictated flavor-suppression of any given diagram. In particular, the least suppressed diagrams for a given process will involve a minimum number of baryon-number violating vertices, and a minimum of flavor changing. For each baryon number violating vertex~(\ref{eqn:BNVrenormW}), all three flavor lines should connect to external quarks; otherwise the diagram involves extra insertions of~(\ref{eqn:BNVrenormW}), and is subdominant.

Thus, we can estimate the amplitude for the diagram by specifying the flavor structure, by which we mean the flavors of the right-handed quarks/squarks connected to the baryon-number violating vertex~(\ref{eqn:BNVrenormW}), as well as the flavors of the external quarks on the flavor lines emanating from the BNV vertex. In addition to the vertex factor, the required charged and/or neutral flavor changes then come with a right $\to$ left Yukawa suppression, together with a CKM suppression for charged flavor-changing or a suppression of the form~(\ref{eqn:FCNCs}) for neutral flavor-changing, whereas quarks/squarks which do not change flavor receive no additional Yukawa suppression.

Given a flavor structure, the heuristic estimation scheme outlined above should give an approximate upper bound on the amplitude, once suppression from the superpartner propagators and loop suppression (if applicable) is accounted for. As the number of possible flavor structures is finite, and much smaller than the number of possible diagrams, it becomes straightforward to obtain an approximate upper bound on the amplitude for all relevant flavor structures.

If we can find a diagram with amplitude equal to the upper bound, then this diagram is probably the dominant contribution to the process in question. The simplest diagrams will often involve only squark flavor-changing, since otherwise additional $W$ bosons are required. In this case, the heuristic scheme outlined above is essentially exact (up to unknown MFV coefficients, which are assumed to be order one). However, if quark flavor changing is involved, the amplitude is somewhat dependent on the details. In particular, while CKM suppression is still present, Yukawa suppression is less obvious. We now consider this point in detail.

For a light quark, the left $\leftrightarrow$ right propagator takes the approximate form $m_q/E^2$,
where $E\sim \Lambda_{QCD} \gg m_q$ is the characteristic energy for the baryon-number violating process. By contrast, for a heavy quark ($m_q \gg E$) the left $\leftrightarrow$ right propagator will take the approximate form $1/m_q$.
In either case, the contribution to the overall amplitude will be made dimensionless by a factor of $\sim E$ in the numerator, arising either from loop integrals or from a hadronic matrix element. Thus, the overall left $\leftrightarrow$ right suppression appears to be only $m_q/E$ and $E/m_q$ for light and heavy quarks respectively, whereas (for light quarks), the assumed Yukawa suppression is much smaller. However, in general left $\leftrightarrow$ right mixing will be followed by charged flavor-changing --- this is the reason for including it in the diagram --- with an associated $g^2/M_W^2 = 2/v^2$ from the $W$ boson propagator, where the dimensions will again be cancelled by factors of $E$. Counting one-half of the $W$ propagator suppression. (the other end of $W$ boson line will lead to flavor changing elsewhere in the diagram), we obtain a net suppression of approximately
\begin{equation}
\frac{m_q}{v/\sqrt{2}} \;\;\; \mathrm{or} \;\;\; \frac{E^2}{m_q\, v/\sqrt{2}} \;\; ,
\end{equation}
for light and heavy quarks, respectively. Thus, for a light quark, the net suppression is the same as Yukawa suppression (for $\tan \beta=1$), whereas for a heavy quark, the diagram is suppressed by an additional factor of $\sim(E/m_q)^2$. At large $\tan \beta$, the suppression is greater than Yukawa suppression for all quarks except for the up-quark, but the difference here is only $1/\sqrt{2}$, and is effectively negligible.

The above argument is more subtle in the case of a loop diagram, since $q^2$ within the loop may be much higher than $\Lambda_{\rm QCD}^2$. Roughly, the net effect is to change the distinction between ``light'' and ``heavy'' quarks; for instance, if $q^2 \sim M_W^2$ within the loop, then only the top quark is ``heavy.'' Yet more subtleties arise for flavor-changing neutral currents of right-handed quarks, since there are then more mass insertions than $W$ vertices. However, the discrepancy is not very important if the mass insertions lie within a loop dominated by loop momentum $q^2 \gsim M_W^2$. Thus, the estimation scheme outlined above also applies qualitatively to quark flavor changing, where the Yukawa suppression now comes partly from $W$ boson propagators and/or loop suppression. Although the exact amplitude will depend on the specifics, this heuristic scheme is a useful way to isolate the larger diagrams contributing to a process of interest.

\section{A systematic search for additional large diagrams} \label{app:search}

We now apply the estimation scheme developed in Appendix~\ref{app:scheme} to search for additional large diagrams which are potentially competitive with those considered in~\S\ref{sec:consmassless} and~\S\ref{sec:consneutrinos}.

\subsection{$n-\bar{n}$ oscillations} \label{subsec:nnbarextra}

We first consider $n-\bar{n}$ oscillations. The amplitude must be built from two insertions of~(\ref{eqn:BNVrenormW}), each of which carries at least one second-generation down-type quark/squark, with all flavor lines connected to external quarks (there are no spectator quarks). As the external quarks are precisely two up-quarks and four down-quarks, the second and third generation quarks must all flavor change to first generation quarks. Furthermore, converting the two squarks into quarks requires the exchange of at least one gaugino or higgsino; any additional three or four-point interactions can only be present at one-loop or higher.

Due to the strong Yukawa suppression of the tree level amplitude~(\ref{eqn:nnbarAmp}), it is conceivable that one-loop amplitudes can be competitive with it. We now search for the largest such diagrams. In any $n-\bar{n}$ oscillation diagram of interest, the external quarks must all be first generation quarks. Thus, for a given flavor structure for the BNV vertex, we can estimate a minimum flavor suppression by assuming charged flavor-changing to first-generation quarks for each leg, since neutral flavor changing is never dominant over charged flavor changing in this context. We estimate the flavor-dependent minimum suppression as
\begin{center}
\renewcommand{\arraystretch}{1.25}
\begin{tabular}{c|ccc}
& $s\, b$ & $b\, d$ & $d\, s$ \\
\hline
$u$ & $y_u y_s^2 y_b^2 \lambda^4/2$ & $y_u y_b^2 y_d\lambda^4/2$ & $y_u y_d y_s^2\lambda^4/2$\\
$c$ & $y_c^2 y_s^2 y_b^2\lambda^6/2$ & $y_c^2 y_b^2 y_d \lambda^4/2$ & $y_c^2 y_d y_s^2\lambda^4$\\
$t$ & $y_s^2 y_b^2\lambda^{10}/2$ & $y_b^2 y_d \lambda^8/2$ & $y_d y_s^2 \lambda^4$
\end{tabular}
\renewcommand{\arraystretch}{1.0}
\end{center}
where the rows and columns correspond to the flavors of the quarks/squarks attached to the BNV vertex.
It is straightforward to check that, for the assumed range $3 \lsim \tan \beta \lsim 45$, $t d s \sim t b d$ gives the weakest suppression, whereas the next weakest, $c b d$, is $\lsim 1/20$ as large.

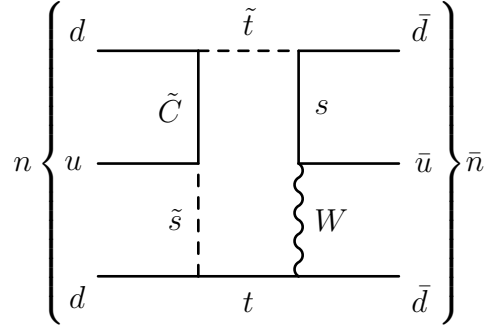
\begin{figure}
\begin{center}
 \begin{fmffile}{nnbaroscillationoneloop}
	        \begin{fmfgraph*}(40,30)
	      \fmfstraight
	       \fmfleft{i1,i2,i3}
	       \fmfright{o1,o2,o3}
	        \fmf{plain}{i1,v1}
                 \fmf{plain,label=$t$}{v1,v2}
                 \fmf{plain}{v2,o1}
	        \fmf{plain}{i2,v3}
                  \fmf{phantom}{v3,v4}
                  \fmf{plain}{v4,o2}
                  \fmf{plain}{i3,v5}
                   \fmf{dashes,label=$\tilde{t}$,label.side=left}{v5,v6}
                  \fmf{plain}{v6,o3}
                  \fmffreeze
                  \fmf{dashes,label=$\tilde{s}$,label.side=left}{v1,v3}
                   \fmf{plain,label=$\tilde{C}$,label.side=left}{v3,v5}
                   \fmf{wiggly,label=$W$}{v2,v4}
                   \fmf{plain,label=$s$}{v4,v6}
 	        \fmflabel{$d$}{i1}
	          \fmflabel{$u$}{i2}
	            \fmflabel{$d$}{i3}
	              \fmflabel{$\bar{d}$}{o1}
	                \fmflabel{$\bar{u}$}{o2}
	                  \fmflabel{$\bar{d}$}{o3}
	        \end{fmfgraph*}
	    \end{fmffile}
\lpar{n}{23mm}{14mm}{-54mm}
\rpar{\bar n}{23mm}{14mm}{1mm}
\end{center}
\caption{The leading one-loop contribution to $n-\bar{n}$ oscillation.\label{fig:nnbarosconeloop}}
\end{figure}

There is only one possible one-loop diagram with two flavor-changing quarks (Fig.~\ref{fig:nnbarosconeloop}). Assuming that the dominant contribution to the loop integral occurs in the range $M_W^2 \lsim q^2 \lsim m_t^2$, we estimate:
\begin{equation} \label{eqn:nnbar1loop2}
\mathcal{M} \sim \frac{g^2}{16 \pi^2} \tilde{\Lambda}\, t_{\beta}^5\, \lambda^8 \frac{m_d^2 m_s^4}{m_t^6} \left( \frac{\tilde{\Lambda}}{m_{\tilde{q}}} \right)^4 \left( \frac{\tilde{\Lambda}}{m_{\chi}} \right) \, ,
\end{equation}
for two $t d s$ vertices, where the $\tan\beta$ dependence is less strong than our naive estimate because the strange-quark left $\leftrightarrow$ right mass insertion is not enhanced at large $\tan \beta$, unlike the corresponding Yukawa coupling. While~(\ref{eqn:nnbar1loop2}) is competitive with~(\ref{eqn:nnbarAmp}) at $\tan\beta=3$, it grows more slowly at large $\tan\beta$, and becomes subdominant. Other combinations of $t d s$ and $t b d$ give a similar result. Since other flavor structures ought to lead to further suppression, we conclude that the tree-level result~(\ref{eqn:nnbarAmp}) is the dominant contribution to $n-\bar{n}$ oscillations at large $\tan\beta$, where the predicted oscillation time is closest to present experimental bounds.

\subsection{Dinucleon decay} \label{subsec:dinucleonextra}

We now consider additional contributions to dinucleon decay. Conservation of electric charge requires that each up-type $\to$ down-type flavor change has a corresponding down-type $\to$ up-type flavor change, which can in principle occur on one of the ``spectator'' flavor lines (those not connected to the BNV vertices). However, each such occurrence is strongly suppressed -- by about $g \Lambda_{\rm QCD}/M_W$ -- due to the $W$ boson propagator, since at most half of the propagator suppression accounts for necessary Yukawa suppression on the ``primary'' flavor lines (those connected to the BNV vertices), as discussed in Appendix~\ref{app:scheme}.

Keeping this suppression in mind, we can search for additional large diagrams by exhaustively cataloging the possible flavor structures for each BNV vertex, grouped together on the basis of the flavors of their external light quarks, estimating the suppression for each flavor structure according to the scheme of Appendix~\ref{app:scheme}. To find the largest diagrams, we find the least suppressed flavor structures for each set of external quarks, and then take the products of all pairs of these suppressions, bearing in mind that for final-state strangeness $|S| \ge 3$, two-body decays are not possible (leading to phase-space suppression), and appending a factor of $\sim g \Lambda_{\rm QCD}/M_W$ for each unit of net charge of the external quarks.

Besides the two diagrams already considered in~\S\ref{subsec:dinucleon}, such a search turns up no flavor structures with a lesser suppression for any $3 \lsim \tan\beta \lsim 45$. Thus we conclude that, to the extent to which the scheme of Appendix~\ref{app:scheme} is valid, the two dominant diagrams are the charged and neutral flavor-changing diagrams already considered.

\subsection{Proton decay} \label{subsec:protonextra}

Finally, we consider additional contributions to proton decay. In the quark sector, we require a single baryon number violating vertex~(\ref{eqn:BNVrenormW}), with a corresponding squark propagator suppression. Requiring that the external quarks be light with strangeness $|\Delta S| \le 1$ and applying the method of Appendix~\ref{app:scheme}, we find that a $t d s$ vertex with $t \to d$ flavor-changing is the least suppressed, with $t \to u$ neutral flavor-changing competitive at large $\tan \beta$. These are the same flavor structure that were considered in \S\ref{sec:consneutrinos}.

However, as argued in~\S\ref{sec:consneutrinos}, the charged-lepton diagram suffers from a chiral suppression. This will occur whenever the squark is up-type and undergoes charged flavor changing, emitting an $\ell^-$ (via mixing with the chargino), i.e. when the net-charge of the external quarks connected to the baryon-number violating vertex is $-1$, since charge conservation otherwise requires the exchange of a $W$ boson with one of the spectator quarks, resulting in a comparable suppression, as disucussed in~\S\ref{subsec:dinucleonextra}. Accounting for the chiral suppression and reapplying the methods of Appendix~\ref{app:scheme}, we conclude that the neutral flavor-changing diagram considered in~\S\ref{sec:consneutrinos} is always dominant.

As the bounds on $|\Delta S| = 0$ decays are somewhat stronger, one might be tempted to consider diagrams of this type. However, according to our estimation scheme, the largest $|\Delta S| = 0$ processes --- $t b d$ with $b \to u, d$ and $t \to d$ flavor changing or $t d s$ with $t \to d$ and $s \to u$ flavor changing --- receive an additional flavor suppression of about $y_s \lambda$, or at least $10^{-2}$ for the assumed range $3 \lsim \tan\beta \lsim 45$. Consequently, $|\Delta S| = 1$ decays are strongly preferred, and their non-observation will lead to the strongest constraints.

\section{Higher dimensional operators} \label{app:highdimension}

We now consider whether higher-dimensional operators can affect our conclusions.
We first consider $|\Delta B |=2$ processes. Lepton-number violating interactions are irrelevant, since they are strongly suppressed by $Y_N$ and $\mu_N = M_N/\Lambda_R$. At dimension five, there is only one allowed baryon-number violating correction, which appears in the K\"{a}hler potential:
\begin{equation} \label{eqn:BNVdim5K}
K_{BNV}^{(5)} = \frac{1}{\Lambda} (Y_u Y_u^{\dag} + Y_d Y_d^{\dag}) Q Q Y_d^{\dag} \bar{d}^{\dag} \, .
\end{equation}
After integrating out the auxiliary fields, this term (combined with the $Q Y_d \bar{d} H_d$ Yukawa coupling), has a similar effect to a $Q^3 H_d$ superpotential term, but with at least two $Y_d$ spurions, leading to a minimum Yukawa suppression of $y_b^2$. Together with the dimension-five $\sim v/\Lambda$ suppression and CKM suppression (of the same form as for~(\ref{eqn:BNVrenormW})), it is straightforward to check that the vertex factor must be substantially smaller than any of those contributing to the dominant diagrams considered in \S\ref{sec:consmassless}\ --- in the latter case we also include any additional suppression from flavor changing --- so long as $\Lambda \gsim 10^{12}\mathrm{\ GeV}$.\footnote{A more detailed analysis might reveal that an even lower cutoff is permissible.} Thus, for a GUT scale cutoff, such contributions are strongly subdominant, whereas dimension six and higher operators are sufficiently suppressed without any flavor suppression.

In the case of nucleon decay, higher-dimensional $|\Delta L|=1$ operators are potentially dangerous. However, they necessarily come with a suppression of at least $\mu_N Y_N^2$ (ignoring flavor structure) in addition to their $\sim v/\Lambda$ cutoff suppression, and are therefore subdominant to the lepton-gaugino mixing induced by the $\mathcal{V}^{(2)}$ spurion. Thus, for a high cutoff, higher dimensional lepton-number violating operators can only be significant if they lead to an enhancement in the quark sector. Specifically, operators which violate lepton \emph{and} baryon number can be dangerous, but these occur first at dimension six, both in the K\"{a}hler potential and the superpotential. Notably, the dangerous (R-parity even) dimension-five operators $Q^3 L$, $\bar{u} \bar{u} \bar{d} \bar{e}$, and $\bar{u} \bar{d} \bar{d} \bar{N}$ are absent from the superpotential due to holomorphy constraints. Dimension six operators are not dangerous in this context, since the smallness of $\mathcal{V}$ spurion (cf.\ (\ref{eqn:Vbound})) combined with cutoff suppression is sufficient to easily evade bounds on the proton lifetime.

\bibliographystyle{JHEP}
\bibliography{mfvsusyfinal}

\end{document}